\documentclass{jaa}
\usepackage{natbib}

\usepackage{graphicx}

\begin{document}\sloppy


\title{New evidence and analysis of cosmological-scale asymmetry in galaxy spin directions}


\author{Lior Shamir\textsuperscript{1,*}}
\affilOne{\textsuperscript{1}Kansas State University, Manhattan KS, USA.\\}

\twocolumn[{

\maketitle

\corres{lshamir@mtu.edu}

\msinfo{1 January 2015}{1 January 2015}

\begin{abstract}

In the past several decades, multiple cosmological theories that are based on the contention that the Universe has a major axis have been proposed. Such theories can be based on the geometry of the Universe, or multiverse theories such as black hole cosmology. The contention of a cosmological-scale axis is supported by certain evidence such as the dipole axis formed by the CMB distribution. Here I study another form of cosmological-scale axis, based on the distribution of the spin direction of spiral galaxies. Data from four different telescopes is analyzed, showing nearly identical axis profiles when the distribution of the redshifts of the galaxies is similar. 

\end{abstract}

\keywords{Galaxy: general -- galaxies: spiral -- cosmology: large-scale structure of Universe -- cosmology: observations.}

}]

\doinum{12.3456/s78910-011-012-3}
\artcitid{\#\#\#\#}
\volnum{000}
\year{0000}
\pgrange{1--}
\setcounter{page}{1}
\lp{1}

\section{Introduction}
\label{introduction}

While the Cosmological Principle has been the common working assumption for most cosmological theories, accumulating evidence suggest that the Universe is not necessarily isotropic. Probes that show cosmological-scale anisotropy include the cosmic microwave background \citep{eriksen2004asymmetries,cline2003does,gordon2004low,campanelli2007cosmic,zhe2015quadrupole}, short gamma ray bursts \citep{meszaros2019oppositeness}, radio sources \citep{ghosh2016probing,tiwari2015dipole,tiwari2016revisiting}, LX-T scaling \citep{migkas2020probing}, Ia supernova \citep{javanmardi2015probing,lin2016significance}, galaxy morphology types \citep{javanmardi2017anisotropy}, dark energy \citep{adhav2011kantowski,adhav2011lrs,perivolaropoulos2014large,colin2019evidence}, polarization of quasars \citep{hutsemekers2005mapping,secrest2021test}, and high-energy cosmic rays \citep{aab2017observation}.


In particular, the anisotropy of the CMB shows certain evidence of an axis of a cosmological scale \citep{abramo2006anomalies,mariano2013cmb,land2005examination,ade2014planck,santos2015influence,dong2015inflation,gruppuso2018evens}. A Hubble-scale axis has been also proposed by the anisotropy in cosmological acceleration rates \citep{perivolaropoulos2014large}.

These observations can be viewed as a shift from the standard cosmological models \citep{pecker1997some,perivolaropoulos2014large,bull2016beyond,velten2020hubble}. Possible explanations include double inflation \citep{feng2003double}, contraction prior to inflation \citep{piao2004suppressing}, primordial anisotropic vacuum pressure \citep{rodrigues2008anisotropic}, moving dark energy \citep{jimenez2007cosmology}, multiple vacua \citep{piao2005possible}, or spinor-driven inflation \citep{bohmer2008cmb}. Other theories can be based on the geometry of the Universe, such as ellipsoidal universe \citep{campanelli2006ellipsoidal,campanelli2007cosmic,campanelli2011cosmic,gruppuso2007complete,cea2014ellipsoidal}, geometric inflation \citep{arciniega2020geometric,edelstein2020aspects,arciniega2020towards,jaime2021viability}, or supersymmetric flows \citep{rajpoot2017supersymmetric}. A cosmological-scale axis is also related to the theory of rotating universe. While early rotating universe theories were based on a static universe \citep{godel1949example}, more recent theories are based on an expanding universe \citep{ozsvath1962finite,ozsvath2001approaches,sivaram2012primordial,chechin2016rotation,seshavatharam2020integrated,camp2021}. In these models, a cosmological-scale axis is expected.

The existence of a cosmological-scale axis can also be linked to holographic big bang \citep{pourhasan2014out,altamirano2017cosmological}, and black hole cosmology \citep{pathria1972universe,easson2001universe,chakrabarty2020toy}. These theories can explain cosmic inflation without the assumption of dark energy. Black holes spin \citep{gammie2004black,mudambi2020estimation,reynolds2021observational}, and the spin of a black hole is inherited from the spin of the star from which it was created \citep{mcclintock2006spin}. If the Universe was formed in a black hole, and the black hole spins, the Universe should have a preferred direction inherited from the spin direction of the black hole \citep{poplawski2010cosmology,seshavatharam2010physics,seshavatharam2020light}, leading to an axis \citep{seshavatharam2020integrated}. Such black hole universe might not be aligned with the cosmological principle \citep{stuckey1994observable}. It has also been shown that holography can impact the entropy of a hierarchy of the large-scale structure \citep{sivaram2013holography}.

This paper analyzes the possibility of a cosmological-scale axis as predicted by some of these models. A spiral galaxy is an extra-galactic object that its visual appearance changes based on the perspective of the observer. The spin directions of galaxies are aligned within cosmic web filaments \citep{kraljic2021sdss}, and alignment in the spin directions of spiral galaxies have been observed among galaxies that are too far from each other to have gravitational interaction  \citep{lee2019galaxy,lee2019mysterious}. Other studies suggested a correlation between the cosmic initial conditions and the spin direction of spiral galaxies, proposing the contention that galaxy spin direction can be used as a probe to study the early Universe \citep{motloch2021observed}. 

The correlation between the location of spiral galaxies and their spin direction cannot be explained by gravity in its standard form, and was therefore defined as ``mysterious'' \citep{lee2019mysterious}. Evidence of non-random distribution has been proposed by numerous previous studies \citep{longo2007cosmic,longo2011detection,shamir2012handedness,shamir2013color,shamir2016asymmetry,shamir2017large,shamir2017photometric,shamir2017colour,shamir2019large,shamir2020pasa,shamir2020patterns,shamir2020large,lee2019galaxy,lee2019mysterious,shamir2021particles,shamir2021large}.  These observations include several different telescopes such as the Sloan Digial Sky Survey \citep{shamir2012handedness,shamir2020patterns,shamir2021particles}, the Panoramic Survey Telescope and Rapid Response System \citep{shamir2020patterns}, the Hubble Space Telescope \citep{shamir2020pasa}, and the Dark Energy Camera \citep{shamir2021large}. The different telescopes show very similar patterns regardless of the telescope being used, and regardless of whether the galaxies images were analyzed manually or automatically \citep{shamir2021large}. 

This study is based on data from four different telescopes -- the Panoramic Survey Telescope and Rapid Response System (PAN-STARRS), the Dark Energy Camera (DECam), the Sloan Digial Sky Survey (SDSS), and the Hubble Space Telescope (HST). The SDSS galaxies also have redshift, which allows to study whether the location of the most likely axis changes with the redshift. Dependence between the location of the most likely axis and the redshift might indicate that the axis, if exists, does not necessarily go through Earth.


\section{Galaxy data}
\label{data}

The primary dataset used in this study is a dataset of spectroscopic objects identified as galaxies from SDSS. However, to test the consistency of the observations across different telescopes, data from three other telescopes was also used. These telescope include Pan-STARRS, DECam, and HST.


\subsection{SDSS data}
\label{sdss}

\begin{figure}[h]
\centering
\includegraphics[scale=0.7]{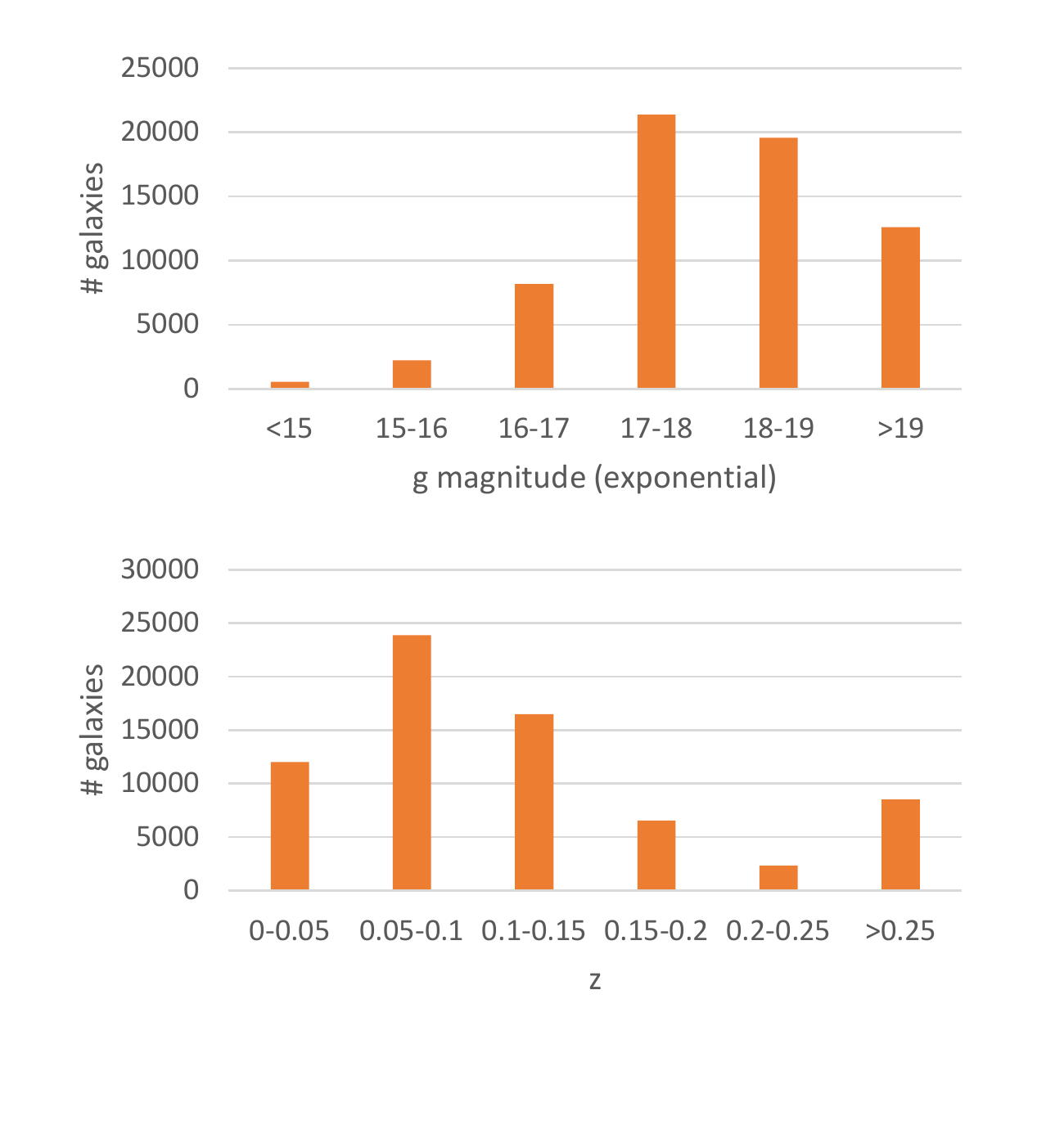}
\caption{Redshift and g magnitude distribution of the SDSS galaxies. All galaxies have spectra.}
\label{distribution_sdss}
\end{figure}

\begin{figure}[h]
\centering
\includegraphics[scale=0.7]{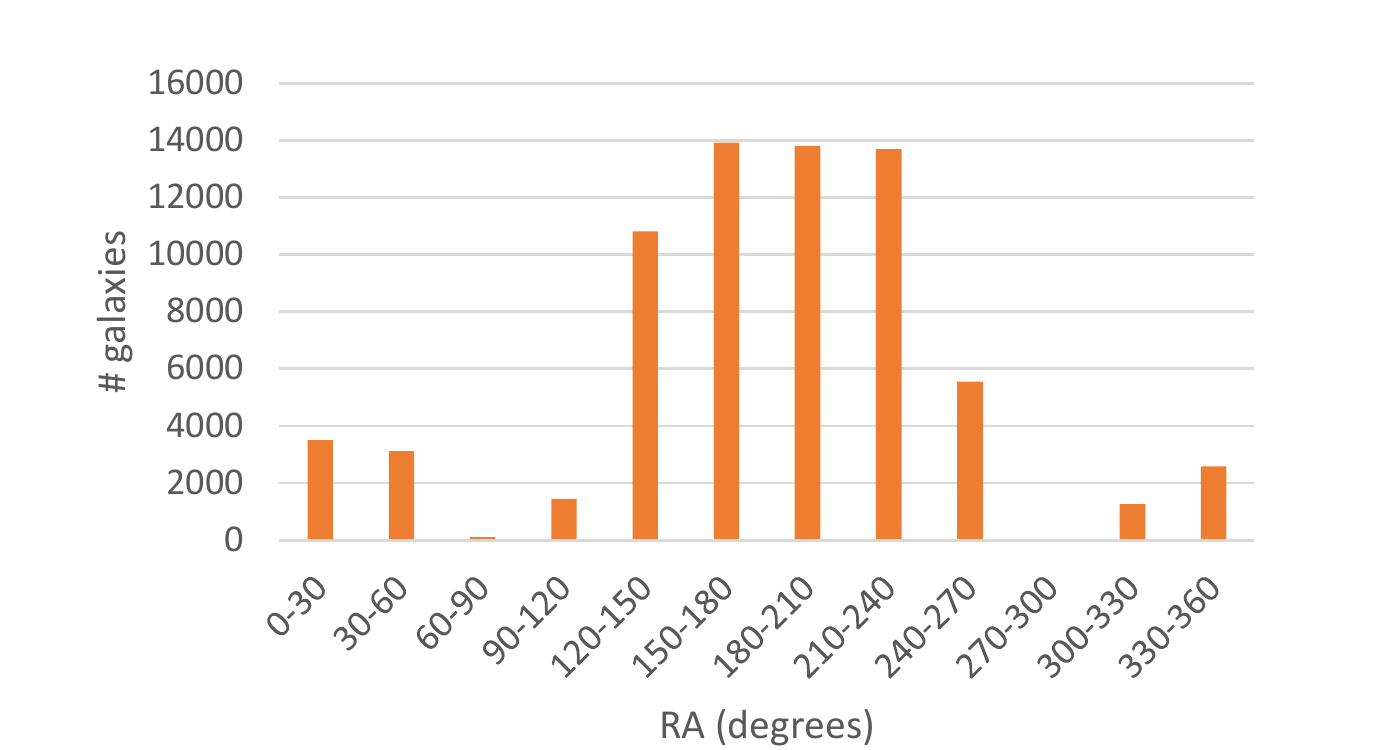}
\caption{Distribution of the SDSS galaxies across different RA ranges.}
\label{distribution_ra_sdss}
\end{figure}

The SDSS galaxies were taken from \citep{shamir2020patterns}. The dataset includes 63,693 galaxies classified with the Ganalyzer algorithm as discussed in Section~\ref{galaxy_classification}. The unique feature of the dataset is that all galaxies have spectra, taken from SDSS DR14. Figure~\ref{distribution_sdss} shows the redshift and g magnitude distribution of the objects in the dataset. Figure~\ref{distribution_ra_sdss} shows the distribution of the galaxy population in different RA ranges. The full details about the dataset are available in \citep{shamir2020patterns}.

\subsection{DECam data}
\label{decam_data}

The dark energy camera (DECam) of the Blanco 4 meter telescope \citep{diehl2012dark,flaugher2015dark} covers $\sim9\cdot10^3$ deg$^2$, mostly from the Southern hemisphere. The DECam image data was retrieved through the DESI Legacy Survey API \citep{dey2019overview}. The description of the dataset and the way it was acquired and annotated is provided in \citep{shamir2021large}. 

In summary, the initial list of objects was taken from Data Release 8 of the DESI Legacy Survey, and included all objects imaged by DECam identified as galaxies, and had magnitude of 19.5 or less in either the g, r or z band. A set of 22,987,246 objects met these criteria. The images of these objects were downloaded by using the {\it cutout} service of the DESI Legacy Survey. Each image is a 256$\times$256 JPEG image, and the Petrosian radius was used to scale the image so that the object fits in the frame. All images were downloaded by the exact same computer, to avoid any differences in the way the files were handled by the system. Downloading such a high number of galaxies naturally required nine months, from June 4th 2020 to March 4th, 2021. The data is described in \citep{shamir2021large}.

As also discussed in Section~\ref{galaxy_classification}, most galaxies are not spiral galaxies or do not have an identifiable spin direction. Therefore, most galaxies were rejected from the analysis and were not assigned with a spin direction. That left a set of 836,451 galaxies in the dataset that were assigned with identifiable spin direction. Some of these galaxies are close satellite galaxies or other large extended objects inside a larger galaxy. To remove such objects, objects that had another object within less than 0.01$^o$ were removed. After removing neighboring objects, 807,898 galaxies were left in the dataset. More information about that dataset is provided in \citep{shamir2021large}.

Table~\ref{decam_ra_distribution} shows the RA distribution of the DECam galaxies. The DECam galaxies do not have redshift, and therefore the distribution of the redshift was determined by using a subset of 17,027 galaxies that are also included in 2dF \citep{cole20052df}. Table~\ref{z_distribution} shows the redshift distribution of the DECam galaxies.  


\begin{table}
\centering
\begin{tabular}{lc}
\hline
RA           & \# galaxies  \\
(degrees) &                  \\
\hline
0-30     &  155,628  \\  
30-60  &  133,683 \\
60-90 &   80,134 \\ 
90-120 & 21,086 \\	
120-150 & 52,842 \\
150-180 &  59,660 \\
180-210 &  58,899 \\
210-240 & 58,112 \\
240-270 & 36,490 \\
270-300 & 2,602 \\
300-330 & 64,869 \\
330-360 & 83,893 \\
\hline
\end{tabular}
\caption{The number of DECam galaxies in different 30$^o$ RA slices.}
\label{decam_ra_distribution}
\end{table}


\begin{table}
\centering
\begin{tabular}{lc}
\hline
z       & \# galaxies  \\
\hline
0-0.05    &  2,089  \\  
0.05-0.1   & 5,487    \\
0.1-0.15   &  4,226  \\ 
0.15 - 0.2  &  1,927 \\	
0.2-0.25 & 784 \\
0.25 - 0.3 &   621 \\
$>$0.3 &  1,893 \\
\hline
\end{tabular}
\caption{The number of DECam galaxies in different redshift ranges. The distribution is determined by a subset of 17,027 galaxies included in the 2dF data release.}
\label{z_distribution}
\end{table}

\subsection{Pan-STARRS data}
\label{pan-starrs}

Pan-STARRS data includes 33,028 galaxies from Pan-STARRS DR1 \citep{shamir2020patterns}. The initial set included 2,394,452 Pan-STARRS objects identified as extended sources by all color bands, and g magnitude brighter than 19 \citep{timmis2017catalog}. These galaxies were annotated automatically as described in Section~\ref{galaxy_classification}. Most objects in the initial list are too small to allow the identification of their morphology, and therefore just 33,028 with identifiable spin patterns were included in the final dataset.

\begin{figure}[h]
\centering
\includegraphics[scale=0.7]{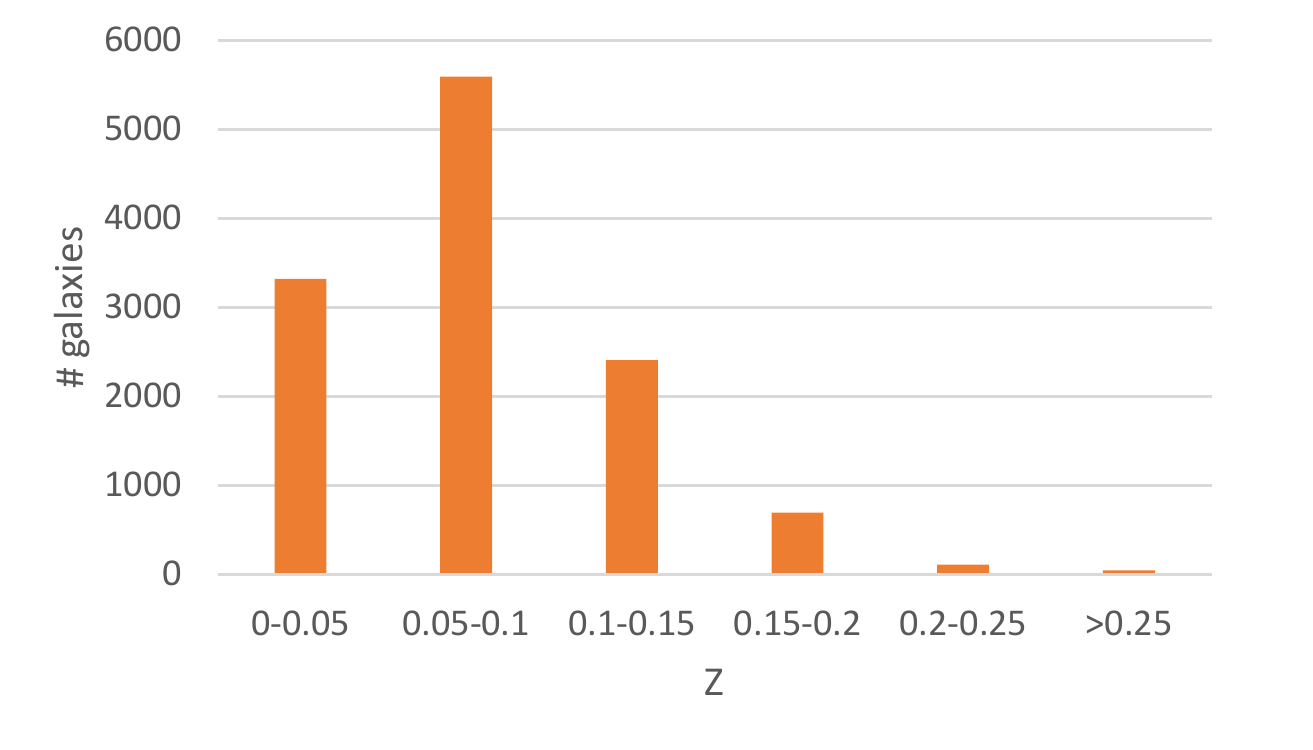}
\caption{The redshift distribution of 12,186 Pan-STARRS galaxies with spectra in SDSS.}
\label{PanSTARRS_z_distribution}
\end{figure}

Figure~\ref{PanSTARRS_z_distribution} shows the distribution of Pan-STARRS galaxies. Since Pan-STARRS galaxies do not have redshift, the distribution was determined by a subset of 12,186 galaxies that had redshift through SDSS.

\subsection{HST data}
\label{hst_data}

While HST cannot provide a dataset of galaxies comparable in size to any of the other digital sky surveys, it has two primary advantages. First of all, the images are not impacted by an atmospheric effect. Although there is no atmospheric effect that can make a galaxy seem to spin in the opposite way, space-based observations can ensure that no unknown atmospheric effect can impact the data. The other advantage is the far deeper field that HST can provide compared to the Earth-based telescope. Because the asymmetry can change with the redshift, using the much more distant HST galaxies can provide information that cannot be provided by the ground-based digital sky surveys.

The dataset of HST galaxies is based on the Hubble Space Telescope (HST) Cosmic Assembly Near-infrared Deep Extragalactic Legacy Survey \citep{grogin2011candels,koekemoer2011candels}. The dataset and the annotation of the galaxies is described in detail in \citep{shamir2020pasa}.

The dataset was taken from several fields: the Cosmic Evolution Survey (COSMOS), the Great Observatories Origins Deep Survey North (GOODS-N), the Great Observatories Origins Deep Survey South (GOODS-S), the Ultra Deep Survey (UDS), and the Extended Groth Strip (EGS). The initial set included 114,529 galaxies \citep{shamir2020pasa}. The image of each galaxy was extracted by using {\it mSubimage} tool \citep{berriman2004montage}, and converted into 122$\times$122 TIF (Tagged Image File) image format.

\begin{figure}[h]
\centering
\includegraphics[scale=0.7]{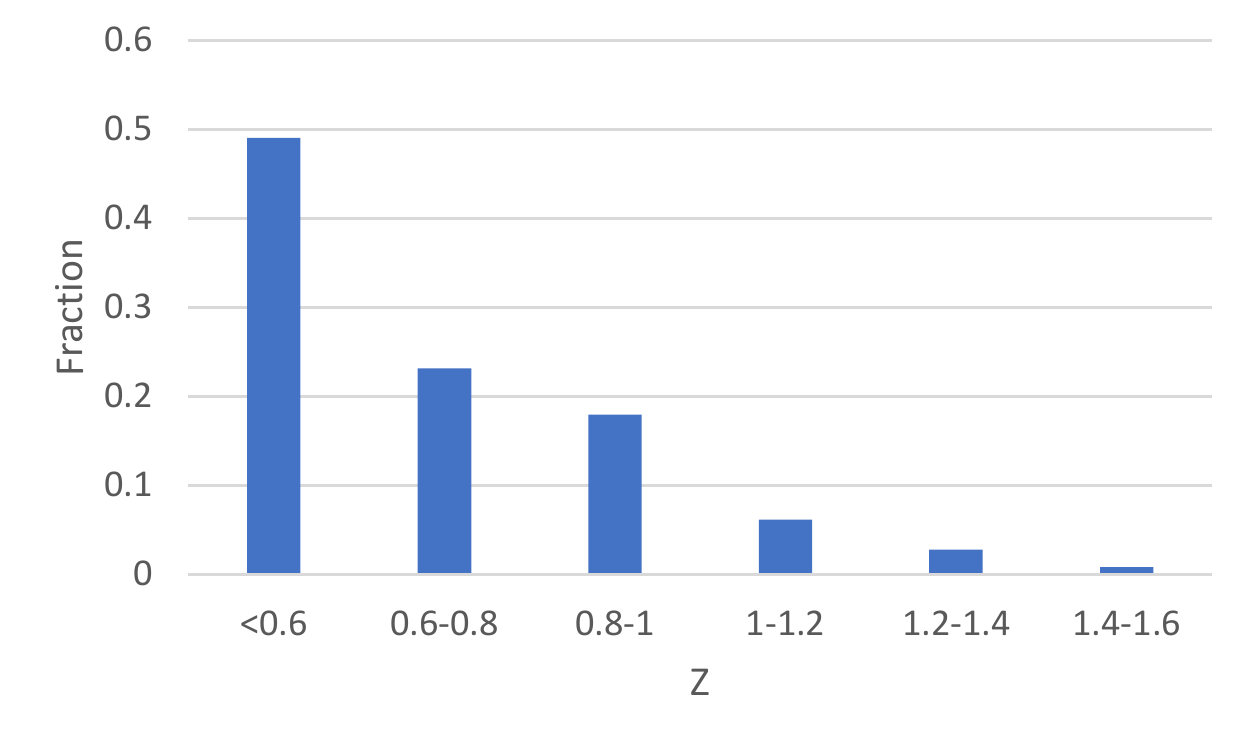}
\caption{The distribution of the photometric redshift of HST galaxies, as a fraction of the total number in each redshift range.}
\label{hst_z}
\end{figure}

Unlike the galaxies of the other telescopes, the HST galaxies were classified by their spin direction manually through a long labor-intensive process. During that process, a random half of the images were mirrored for the first cycle of annotation, and then all images were mirrored for a second cycle of annotation as described in \citep{shamir2020pasa} to offset a possible perceptional bias. That process led to a clean dataset of 8,690 galaxies \citep{shamir2020pasa}. Naturally, the HST galaxies are more distant than the galaxies imaged by the other telescopes. Figure~\ref{hst_z} shows a histogram of the distribution of the photometric redshift of the HST galaxies. The photometric redshift is highly inaccurate, but even when considering the expected inaccuracy of the photometric redshift it can be safely assumed that the HST galaxies are more distant from Earth compared to the galaxies imaged by the ground-based telescopes. Table~\ref{CANDELS} shows the number of galaxies taken from each field.

\begin{table}
\caption{The number of galaxies in each of the five HST fields.}
\label{CANDELS}
\small
\centering
\begin{tabular}{lccc}
\hline
Field & Field                        & \# galaxies   & Annotated \\ 
        & center                     &                    & galaxies    \\ 
\hline
COSMOS &  $150.12^o,2.2^o$ & 84,424 & 6,081 \\ 
GOODS-N & $189.23^o,62.24^o$ & 5,931 & 769 \\ 
GOODS-S & $53.12^o,-27.81^o$  & 5,024 & 540 \\ 
UDS        &  $214.82^o,52.82^o$   & 14,245 &  616 \\ 
EGS        &  $34.41^o,-5.2^o$    & 4,905    &  684 \\ 
\hline
\end{tabular}
\end{table}


\subsection{Distribution of the galaxy population}
\label{galaxy_distribution}

The HST galaxies are concentrated in five relatively small fields, making the distribution of these galaxies simple. The other sky surveys cover far larger footprints. Although the footprints are large, they cover merely part of the sky. Moreover, the density of the galaxy population within the footprint of each sky survey is not uniform. Table~\ref{decam_ra_distribution} shows the distribution of the galaxies in different RA ranges in DECam, and Figure~\ref{distribution_ra_sdss} displays the distribution of the galaxy population in SDSS, showing that the distribution is not uniform.

\begin{figure}[h]
\centering
\includegraphics[scale=0.25]{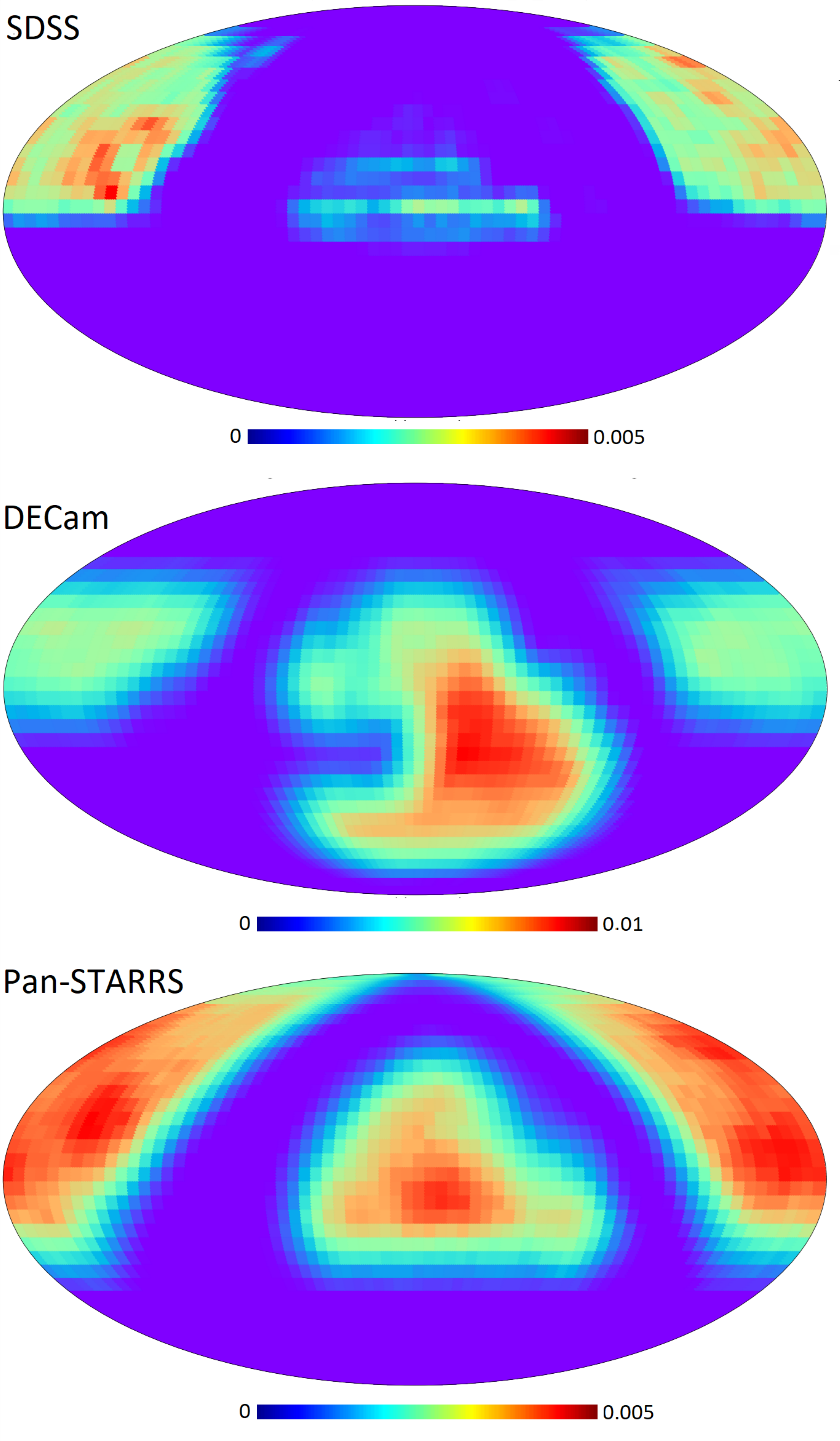}
\caption{The density of the galaxy population in SDSS, Pan-STARRS and DECam. The galaxy population density in each $5^o\times5^o$ field of the sky is determined by the number of galaxies in that $5^o\times5^o$ field divided by the total number of galaxies from that telescope.}
\label{population}
\end{figure}

Figure~\ref{population} shows the galaxy population density in each part of the sky in each of the ground-based telescopes. The distribution is determined by the number of galaxies in each $5^o\times5^o$ field of the sky. To avoid bias due to the different surface area covered by pixels closer to the celestial poles, the Hierarchical Equal Area isoLatitude Pixelization (HEALpix) was used. That analysis shows the differences in the density of galaxies in different parts of the sky and in different telescopes. As expected, the different sky surveys have different footprints, and different galaxy distribution within their footprints.

\begin{figure}[h]
\centering
\includegraphics[scale=0.7]{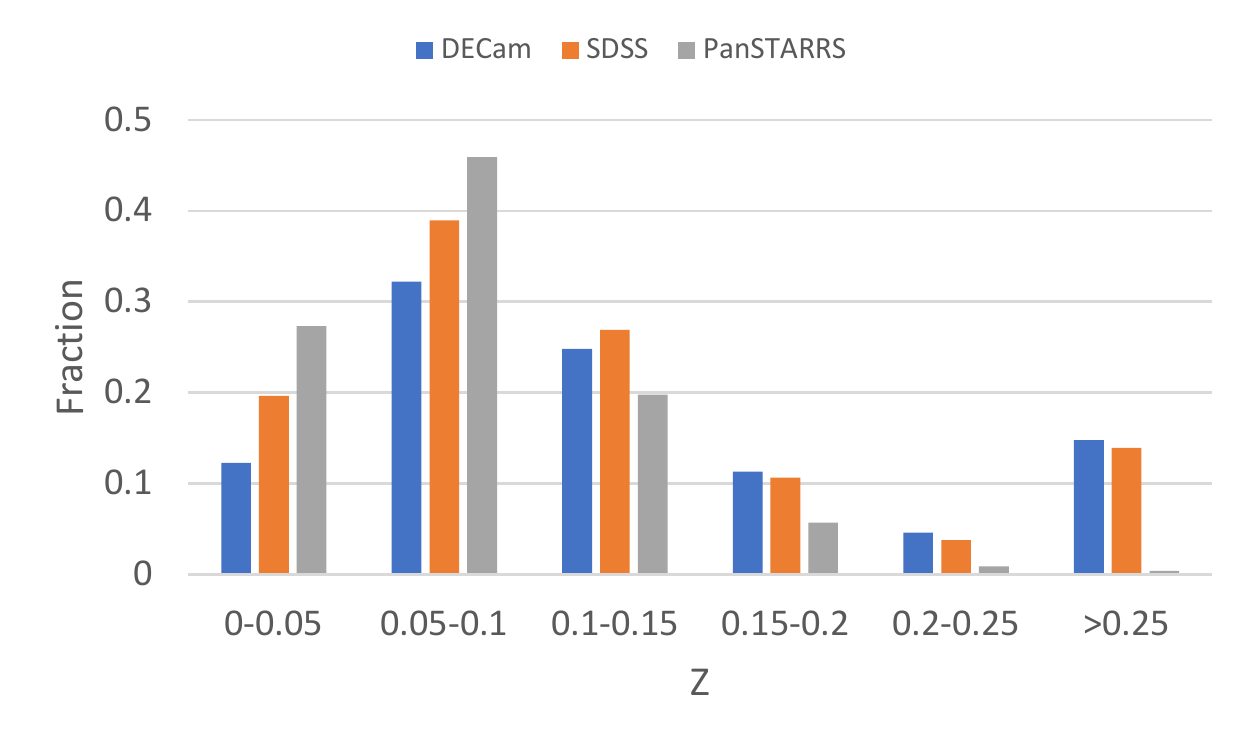}
\caption{The distribution of the redshift in SDSS, DECam, and Pan-STARRS.}
\label{compare_z}
\end{figure}

The different telescopes are not different just in the parts of the sky they cover, but also in the redshift of the galaxies they can image. Obviously, the space-based HST is far deeper than any existing ground-based telescope, and the redshift distribution as shown in Figure~\ref{hst_z} is not comparable to any of the ground-based telescopes. The redshift distribution of the three other telescopes is shown in Figure~\ref{compare_z}.

\subsection{Galaxy annotation}
\label{galaxy_classification}

The HST galaxies were annotated manually, as explained in Section~\ref{hst_data} and in \citep{shamir2020pasa}. The other datasets were far too large to analyze manually, and therefore automatic analysis was needed.

In the recent years, pattern recognition, and especially deep convolutional neural networks (DCNNs), have been becoming very common in classification of galaxy images \citep{sharma2018galaxy,ciprijanovic2020deepmerge}. While DCNNs are powerful tools that provide strong classification accuracy for supervised machine learning tasks, they are based on non-intuitive data-driven rules that are difficult to conceptualize. Because these rules are complex and non-intuitive, it is very difficult to ensure their symmetricity. Even a seemingly unimportant aspect such as the order of the training images can lead to a different DCNN and impact the way it works. As shown in \citep{shamir2021particles}, even subtle bias in the classification algorithm can lead to very strong asymmetry in the analysis. Since the asymmetry or other biases in DCNNs are difficult to control or profile, these algorithms are not suitable for identifying subtle asymmetries. 

To analyze the galaxy images in a controlled and fully symmetric annotation, the Ganalyzer algorithm was used \citep{shamir2011ganalyzer}. Ganalyzer is not based on any aspect of machine learning, deep learning, or pattern recognition. Instead, it is a model-driven algorithm that follows mathematically defined rules. Ganalyzer first transforms each galaxy image into its radial intensity plot, which is a 360$\times$35 image that reflects the changes of pixel intensity relative to the center of the image. Each pixel $(x,y)$ in the radial intensity plot is the median value of the 5$\times$5 pixels around coordinates $(O_x+\sin(\theta) \cdot r,O_y-\cos(\theta)\cdot r)$ in the original galaxy image, where {\it r} is the radial distance measured in percent of the galaxy radius, $\theta$ is the polar angle, and $(O_x,O_y)$ are the pixel coordinates of the center of the galaxy.

Pixels on the galaxy arms are expected to be brighter than pixels that are not on the arm of the galaxy at the same radial distance from the galaxy center. Therefore, peaks in the radial intensity plot are expected to correspond to pixels on the arms of the galaxy at different distances from the galaxy center. A peak detection algorithm \citep{morhavc2000identification} applied to the lines in the radial intensity plot allows to identify the peaks. Linear regression is then applied to the peaks in adjunct lines. The slope of the lines formed by the peaks reflects the curves of the arms, and consequently the spin direction of the galaxy.

\begin{figure}[h]
\centering
\includegraphics[scale=0.5]{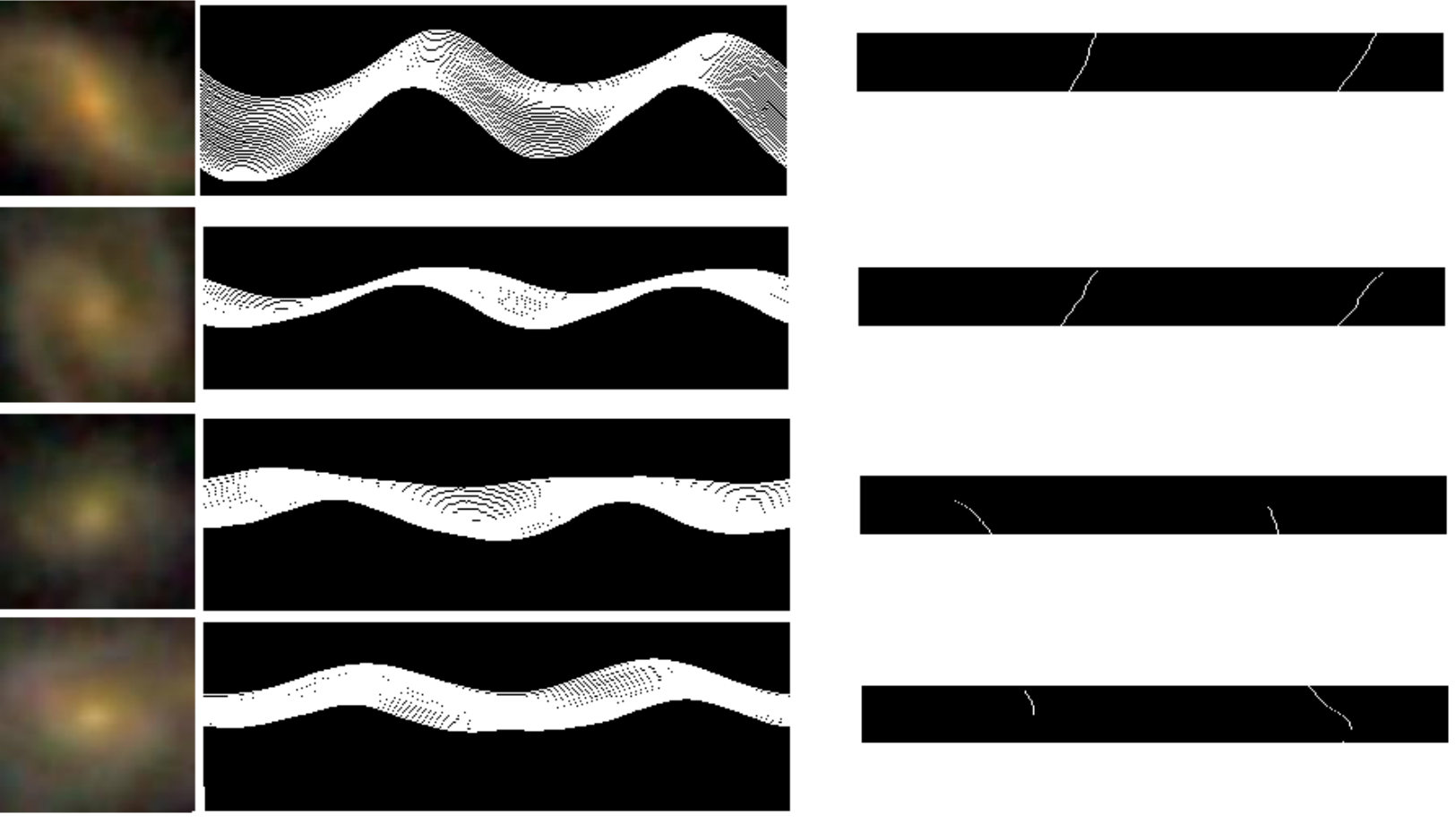}
\caption{Galaxy images, their radial intensity plots, and the peaks of the radial intensity plots of four different SDSS galaxies.}
\label{radial_intensity_plots}
\end{figure}

Figure~\ref{radial_intensity_plots} shows examples of galaxy images, their radial intensity plots, and the peaks identified in the radial intensity plots. As the figure shows, each arm is reflected by a vertical line of peaks. The slope of that line indicates the curve of the arms, and consequently the spin directions of the galaxies. More information about Ganalyzer and detailed performance analysis can be found in \citep{shamir2011ganalyzer,dojcsak2014quantitative,shamir2017photometric,shamir2017colour,shamir2017large,shamir2019large,shamir2020patterns,shamir2021large}.

Since not all galaxies are spiral galaxies, and since not all spiral galaxies have an identifiable spin direction, it is clear that the majority of the galaxies cannot be used for the analysis. Therefore, only galaxies that have at least 30 identified peaks in the radial intensity plot aligned at the same direction are used. Galaxies that do not meet that criterion are rejected regardless of the sign of the linear regression of their peaks. As mentioned above, the primary feature of the algorithm is that it is fully symmetric, and works by clear defined rules.

\section{Analysis method}
\label{analysis_method}

A dipole axis in the distribution of spin directions of galaxies means that the number of spiral galaxies spinning clockwise is higher than the number of galaxies spinning counterclockwise in one hemisphere, but lower in the opposite hemisphere. A simple method  is to separate the sky into two hemispheres, and count the number of clockwise and counterclockwise galaxies in each part of the sky. A simple binomial distribution can then provide the statistical significance of the asymmetry. 

The advantage of the method is that it is simple, and allows a straightforward statistical analysis. An inverse asymmetry in the two opposite hemispheres indicates on possible asymmetry, but also shows that the asymmetry cannot be driven by a bias in the annotation, as such bias is expected to be consistent throughout the entire sky. That will be discussed thoroughly in Section~\ref{error} The downside of this approach is that none of the sky surveys covers the entire sky. Also, the density of the population of the galaxies imaged by each sky survey changes in different parts of the sky, and is not consistent or equally distributed. Ground-based sky surveys can only cover one hemisphere, but in practice cover just a portion of the hemisphere they can observe. That can naturally lead to substantial inconsistencies between different telescopes, as different sky surveys that cover different parts of the sky.

To analyze a dipole axis in a manner that can handle different parts of the sky, a possible axis in the galaxy spin directions of the galaxies can be determined by applying $\chi^2$ statistics to fit the spin direction distribution to cosine dependence, as was done in \citep{shamir2012handedness,shamir2019large,shamir2020pasa,shamir2020patterns}. From each possible $(\alpha,\delta)$ combination in the sky, the angular distance $\phi_i$ between $(\alpha,\delta)$ and each galaxy {\it i} in the dataset was computed. The $\chi^2$ from each $(\alpha,\delta)$ was determined by Equation~\ref{chi2}
\begin{equation}
\chi^2_{\alpha,\delta}=\Sigma_i \frac{(d_i \cdot | \cos(\phi_i)| - \cos(\phi_i))^2}{\cos(\phi_i)} ,
\label{chi2}
\end{equation}
where $d_i$ is the spin direction of galaxy {\it i} such that $d_i$ is 1 if the galaxy {\it i} spins clockwise, and -1 if the galaxy {\it i} spins counterclockwise. 

The $\chi^2$ is computed using the actual spin directions of the galaxies. Then, the $\chi^2$ computed with the actual spin directions is compared to the average $\chi^2$ computed in $10^3$ runs such that $d_i$ of each galaxy {\it i} is assigned with a random number within \{-1,1\}. The standard deviation of the $\chi^2$ of the $10^3$ runs is also computed. Then, the $\sigma$ difference between the $\chi^2$ computed with the actual spin directions and the $\chi^2$ computed with the random spin directions provided the probability to have a dipole axis in that $(\alpha,\delta)$ combination by chance.

Repeating that process from each possible $(\alpha,\delta)$ provides a full map of likelihood to have a dipole axis at each point in the sky. Mollweide projection is then used with the Matplotlib library to visualize the results. As will be discussed in Section 5, the measurement used in this study is a relative measurement rather than an absolute measurement. That is, the measurement is not based on differences in the number of galaxies in different parts of the sky. Instead, the measurement is based on the ratio between galaxies with opposite spin directions. Because the number of galaxies spinning clockwise in any field is expected to be within the same order of magnitude of the number of galaxies spinning counterclockwise, non-uniform distribution of the number of galaxies in the sky cannot lead to signal unless the ratio between clockwise and counterclockwise galaxies is indeed statistically different than 1:1. Detailed explanation and experimental results are described in Section~\ref{uneven_distribution}. 

Since the analysis attempts to find the best fit into cosine dependence, and since the measurement is relative and not absolute, it is not dependent on the part of the sky being covered, or on the distribution of the galaxy population in the dataset. Because the measurement is relative and not absolute, it is also allows to avoid the assumption of uniform sky coverage, as done in analysis of probes such as CMB. For instance, the obstruction of the Milky Way is expected to affect the number of clockwise galaxies in the same way it affects the number counterclockwise galaxies in the potentially obstructed field, and therefore the effect on both types of galaxies is offset. The analysis method is also described in \citep{shamir2012handedness,shamir2019large,shamir2020pasa,shamir2020patterns,shamir2021particles}.

\section{Results}
\label{results}

As a very simple first experiment, the sky was divided into two simple hemispheres by the RA. The first hemisphere was the RA range of $(0^o-180^o)$, and opposite hemisphere $(180^o-360^o)$. The number of galaxies spinning clockwise and the number of galaxies spinning counterclockwise can be compared. Tables~\ref{hemispheres_decam} and~\ref{hemispheres_SDSS} show the number of galaxies spinning clockwise and counterclockwise in the two hemispheres in DECam and SDSS, respectively.

As Table~\ref{hemispheres_decam} shows, DECam have an excessive number of galaxies spinning clockwise in the $(0^o-180^o)$ hemisphere, and more galaxies spinning counterclockwise in the opposite $(180^o-360^o)$ hemisphere. In both cases the differences are statistically significant. DECam mostly covers the Southern hemisphere. In SDSS, which covers mostly the Northern hemisphere, the asymmetry is inverse. In SDSS a higher number counterclockwise galaxies is observed in the $(0^o-180^o)$ hemisphere, and more clockwise galaxies in the opposite hemisphere. These differences are also statistically significant according to binomial distribution when assuming that the probability of a galaxy to spin clockwise or counterclockwise is 0.5.

\begin{table}
\centering
\small
\begin{tabular}{lcccc}
\hline
Hemisphere       & \# cw    & \# ccw    & $\frac{cw-ccw}{cw+ccw}$  & P \\
                 & galaxies & galaxies  &                          &   \\
\hline
$(0^o-180^o)$    &   252,478    & 250,555  &   0.0038      &  0.0033  \\   
$(180^o-360^o)$  &   151,948    & 152,917  &   -0.0033    &  0.039   \\
\hline
\end{tabular}
\caption{Number of clockwise and counterclockwise galaxies in opposite hemispheres in DECam. The P values are the binomial distribution probability to have such difference or stronger by chance when assuming 0.5 probability for a galaxy to spin clockwise or counterclockwise.}
\label{hemispheres_decam}
\end{table}

\begin{table}
\centering
\small
\begin{tabular}{lcccc}
\hline
Hemisphere       & \# cw    & \# ccw    & $\frac{cw-ccw}{cw+ccw}$  & P \\
                 & galaxies & galaxies  &                          &   \\
\hline
$(0^o-180^o)$    &   14,403    & 15,101 &   -0.024      &  0.00002  \\   
$(180^o-360^o)$  &   17,263    & 16,926   & 0.01      &  0.035   \\
\hline
\end{tabular}
\caption{Number of clockwise and counterclockwise galaxies in opposite hemispheres in SDSS.}
\label{hemispheres_SDSS}
\end{table}

\begin{figure}[h]
\centering
\includegraphics[scale=0.15]{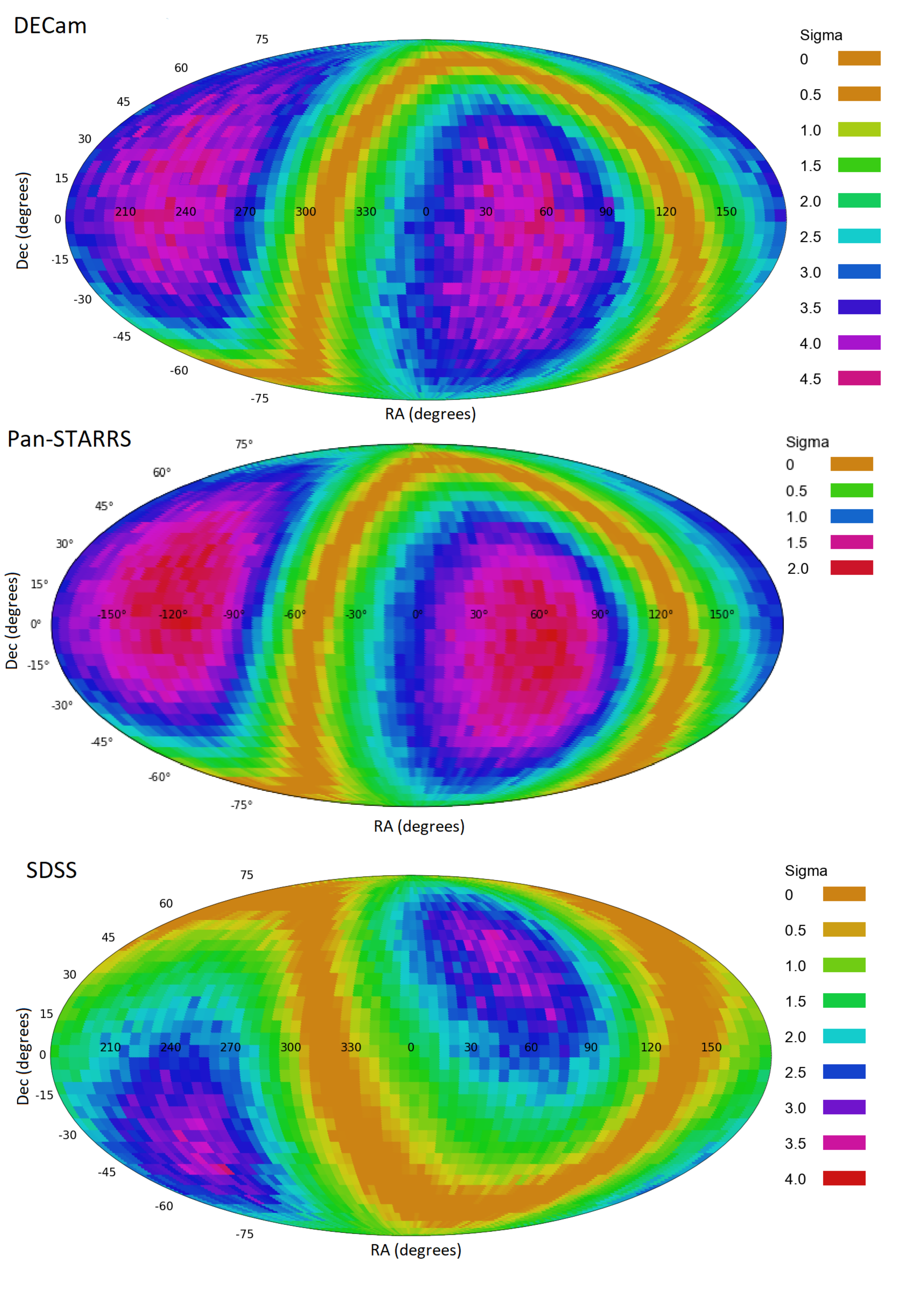}
\caption{Mollweide  projection of the probability of a dipole axis in galaxy spin directions from different $(\alpha,\delta)$ combinations in SDSS, Pan-STARRS, and DECam.}
\label{dipole_datasets}
\end{figure}

While the separation of the sky into two hemispheres is very simple, it relies on a basic and straightforward statistical analysis. To better profile the existence of a dipole axis in the distribution of spin directions of spiral galaxies, the method described in Section 3 was applied. Figure~\ref{dipole_datasets} shows a Mollweide  projection of the probability of a dipole axis in each integer $(\alpha,\delta)$ combination in the datasets acquired from SDSS, Pan-STARRS, and DECam as described in Section~\ref{data}.

As the figure shows, DECam and Pan-STARRS show a nearly identical profile of asymmetry. The most likely axis in DECam data is identified at $(\alpha=57^{+35^o}_{ -35^o} ,\delta=-10^{+66^o}_{-29^o})$, with statistical signal of $4.66\sigma$ \citep{shamir2021large}. 
That axis is very close to the most likely axis identified with the Pan-STARRS data, at $(\alpha=47^o,\delta=-1^o)$, with statistical significance of 1.87$\sigma$ \citep{shamir2020patterns}. It is interesting that the axis is close to the CMB cold spot at $(\alpha=49,\delta=-19^o)$. 

\begin{figure}[h]
\centering
\includegraphics[scale=0.25]{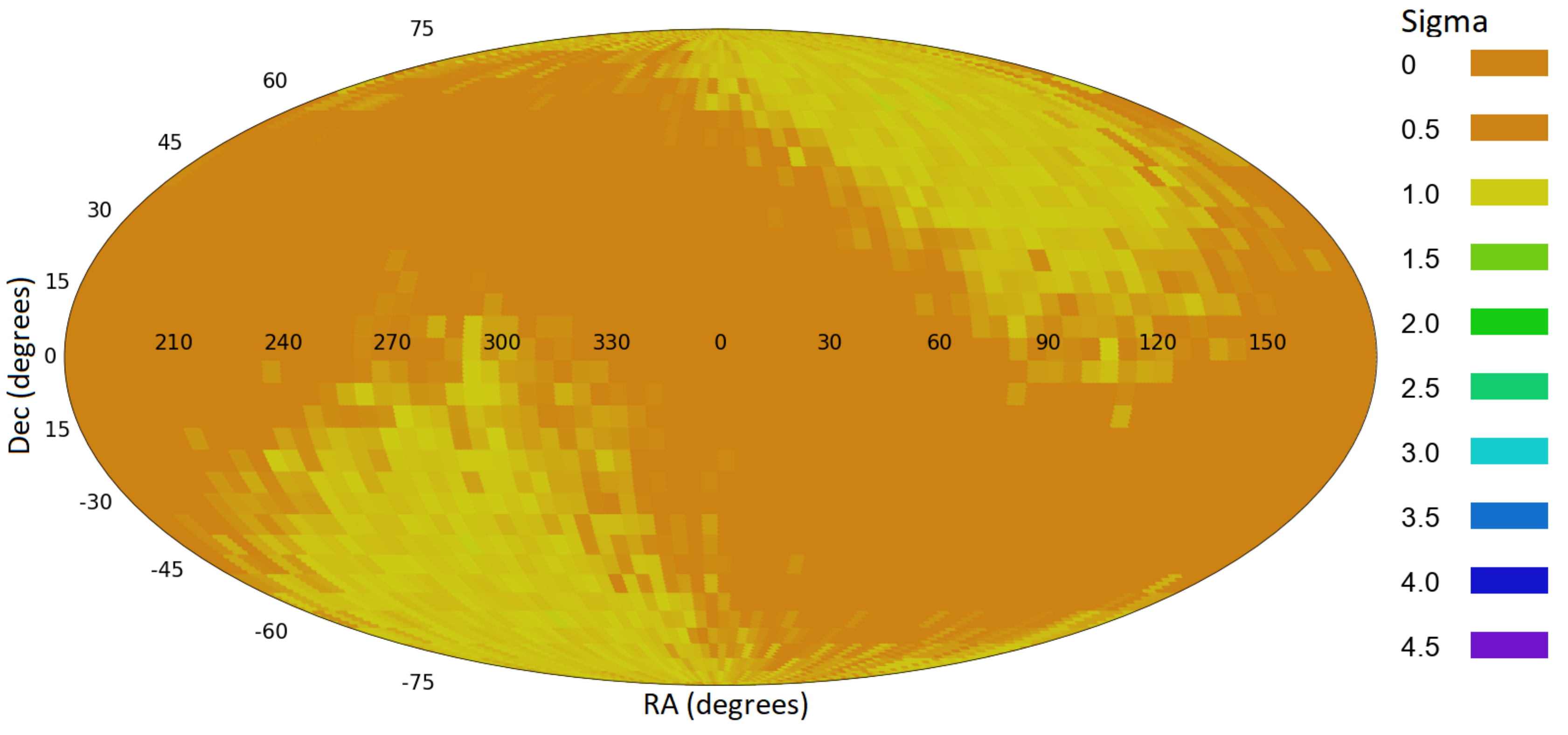}
\caption{The probability of a dipole axis in galaxy spin directions in DECam data such that the galaxies are assigned with random spin directions.}
\label{decam_dipole_random}
\end{figure}

When assigning the galaxies with random spin directions, the signal is immediately lost \citep{shamir2021particles}, and no profile can be identified in any of the telescopes \citep{shamir2020patterns, shamir2021large}. For instance, Figure~\ref{decam_dipole_random} shows the profile of asymmetry in the DECam galaxies such that each galaxy was assigned with a random spin direction. A more detailed analysis of ``sanity tests'' will be discussed in Section~\ref{error}.

The analysis done with the SDSS galaxies shows an axis that is close but not identical to DECam and Pan-STARRS. The axis peaks at $(\alpha=69^{+38^o}_{ -50^o} ,\delta=56^{+21^o}_{-31^o})$, with statistical significance of 4.63$\sigma$.  
While that axis is not identical to the axis identified with the DECam data, it is still within 1$\sigma$ from it. While the axes are still within 1$\sigma$ from each other, the SDSS galaxies have different redshift distribution compared to DECam. For instance, 14.7\% of the DECam galaxies have redshift greater than 0.25, while in SDSS just 13.5\% of the galaxies are in the redshift range. To make a better comparison between the sky surveys, the SDSS galaxy dataset was reduced to a dataset of 38,264 such that the distribution of the redshifts of the galaxies fits the distribution of the redshift of the DECam galaxies. Since all SDSS galaxies have redshift, the normalization of the dataset was done by adding galaxies into the different redshift bins in random order, and excluding a galaxy if the fraction of the galaxies out of the total number of galaxies in its redshift bin is greater than the fraction of that bin in the DECam galaxies. That process leads to a dataset that might be smaller than the original dataset, but its redshift distribution is similar to the dataset it is compared with. Obviously, the process is only possible in datasets were all galaxies have spectra. More information about that process is provided in \citep{shamir2020patterns}. 

\begin{figure}[h]
\centering
\includegraphics[scale=0.25]{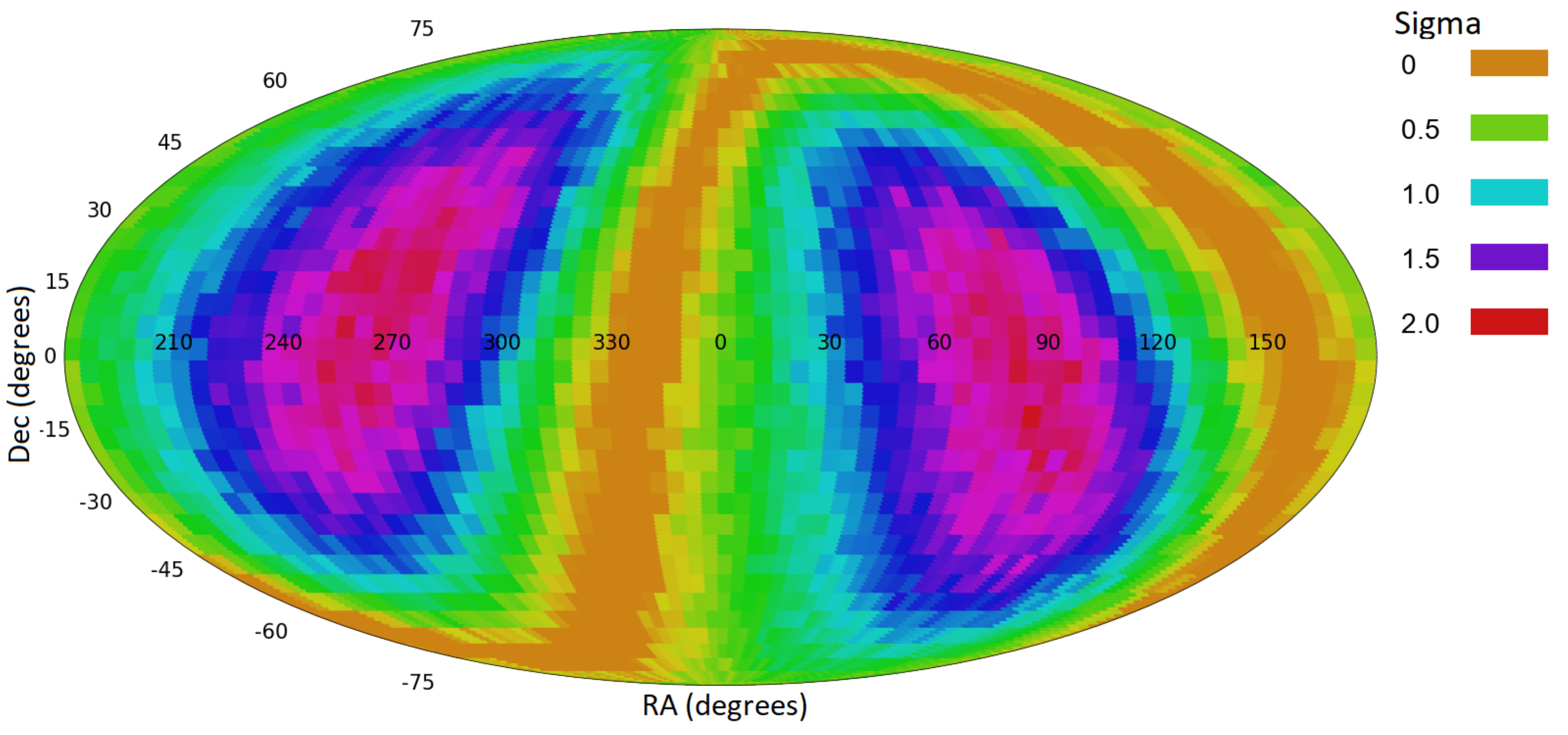}
\caption{The probability of a dipole axis in galaxy spin directions from different $(\alpha,\delta)$ combinations in SDSS when the redshift distribution is similar to the redshift distribution of the DECam galaxies.}
\label{dipole_sdss}
\end{figure}

Figure~\ref{dipole_sdss} shows the probabilities of a dipole axis from the different $(\alpha,\delta)$ combinations. As the figure shows, when normalizing the distribution of the redshift of the SDSS galaxies, the profile is very similar to the profile of the DECam galaxies. The most likely axis identified in the dataset normalized $(\alpha=78^o,\delta=-12^o)$, close to the most likely axis identified in DECam and Pan-STARRS data. In any case, these axes are all within the 1$\sigma$ error, and show good agreement between the different telescopes.


\subsection{Analyzing the impact of the redshift distribution}
\label{redshift_impact}

\begin{figure}[h]
\centering
\includegraphics[scale=0.45]{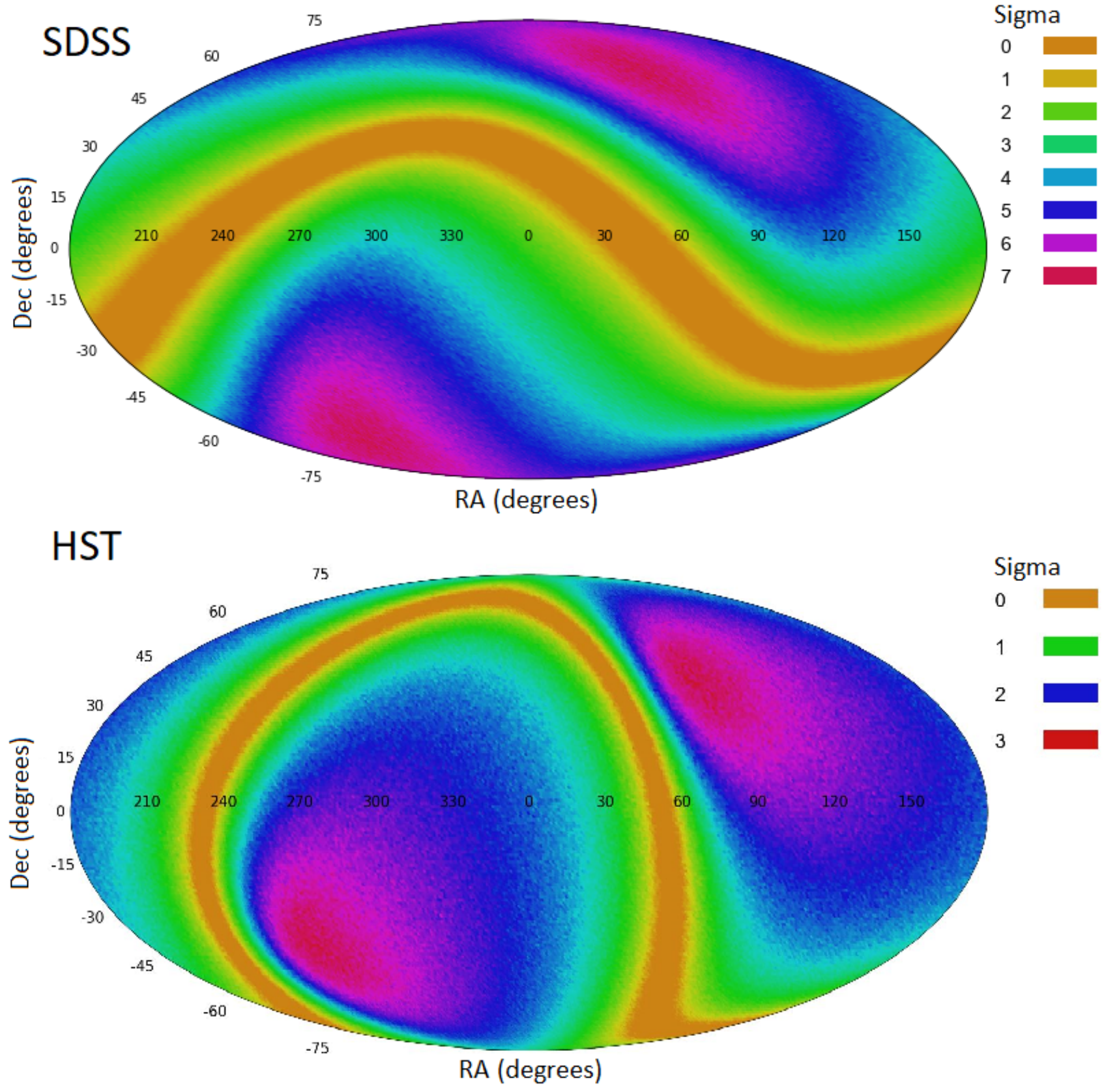}
\caption{The probability of a dipole axis in galaxy spin directions from different integer $(\alpha,\delta)$ combinations in HST and SDSS when the galaxies are limitied to z$>$0.15.}
\label{sdss_hst}
\end{figure}

As discussed above, the redshift of the galaxies have an impact on the location of the most likely axis. As a space-based instrument, HST can image deeper objects than the ground-based telescopes. The mean redshift of HST galaxies as approximated by the photometric redshift is 0.58 \citep{shamir2020pasa}. Although the photometric redshift is highly inaccurate, it can be safely assumed that the HST galaxies have higher redshift than the SDSS galaxies. The dipole axis formed by HST galaxies was compared to the dipole axis formed by the 15,863 SDSS galaxies with $z>0.15$. Figures~\ref{sdss_hst} shows the probabilities of the axes from each possible integer $(\alpha,\delta)$ combination. 

As the figure shows, when the redshift of the galaxies is high, the axes observed in the two telescopes are in close agreement. The most likely dipole axis identified in the HST galaxies is at $(\alpha=78^o,\delta=47^o)$. That is in very close proximity to the most likely dipole axis in the SDSS galaxies with $z>0.15$, peaking at $(\alpha=71^o,\delta=61^o)$ as also shown in \citep{shamir2020pasa}. The statistical signal of the axis observed with HST galaxies is 2.8$\sigma$, and the SDSS axis probability is 7.4$\sigma$. That difference in the statistical significance can be explained by the fact that the HST galaxies are not distributed in the sky as the SDSS galaxies, but are concentrated in five fields. These five fields might not be located where the asymmetry necessarily peaks. For instance, the most populated HST field is COSMOS, centered around $(\alpha=150^o,\delta=2^o)$, where according to the SDSS galaxies the asymmetry exists but does not peak. Table~\ref{cosmos_sdss} shows the distribution of the galaxies in the COSMOS field of HST, and in the $10^o\times10^o$ centered around COSMOS in the other telescopes. All fields show the same direction of the asymmetry, but the asymmetry in the populated fields of HST and DECam is just $\sim$2\%.

\begin{table}
\centering
\scriptsize
\begin{tabular}{lccc}
\hline
Sky          &        \# Clockwise   & \# Counterclockwise & P \\
survey      &        galaxies          & galaxies                   & value \\
\hline
COSMOS (HST) &  3,116 & 2,965 & 0.027 \\
Pan-STARRS     &  222  & 190 & 0.06 \\
SDSS  &  461  & 440 & 0.24 \\
DECam  & 2,640  & 2,498 & 0.025 \\
\hline
\end{tabular}
\caption{The number of clockwise and counterclockwise galaxies in the HST COSMOS field, and in the $10^o\times10^o$ field of SDSS, Pan-STARRS, and DECam centered at COSMOS. The P values are the one-tail binomial probabilities of having asymmetry equal or greater than the observed asymmetry when the probability of a galaxy to spin in a certain direction is 0.5.}
\label{cosmos_sdss}

\end{table}

\subsection{Change of the axis as a function of the redshift}
\label{redshift_change}

\begin{figure*}[h]
\centering
\includegraphics[scale=0.215]{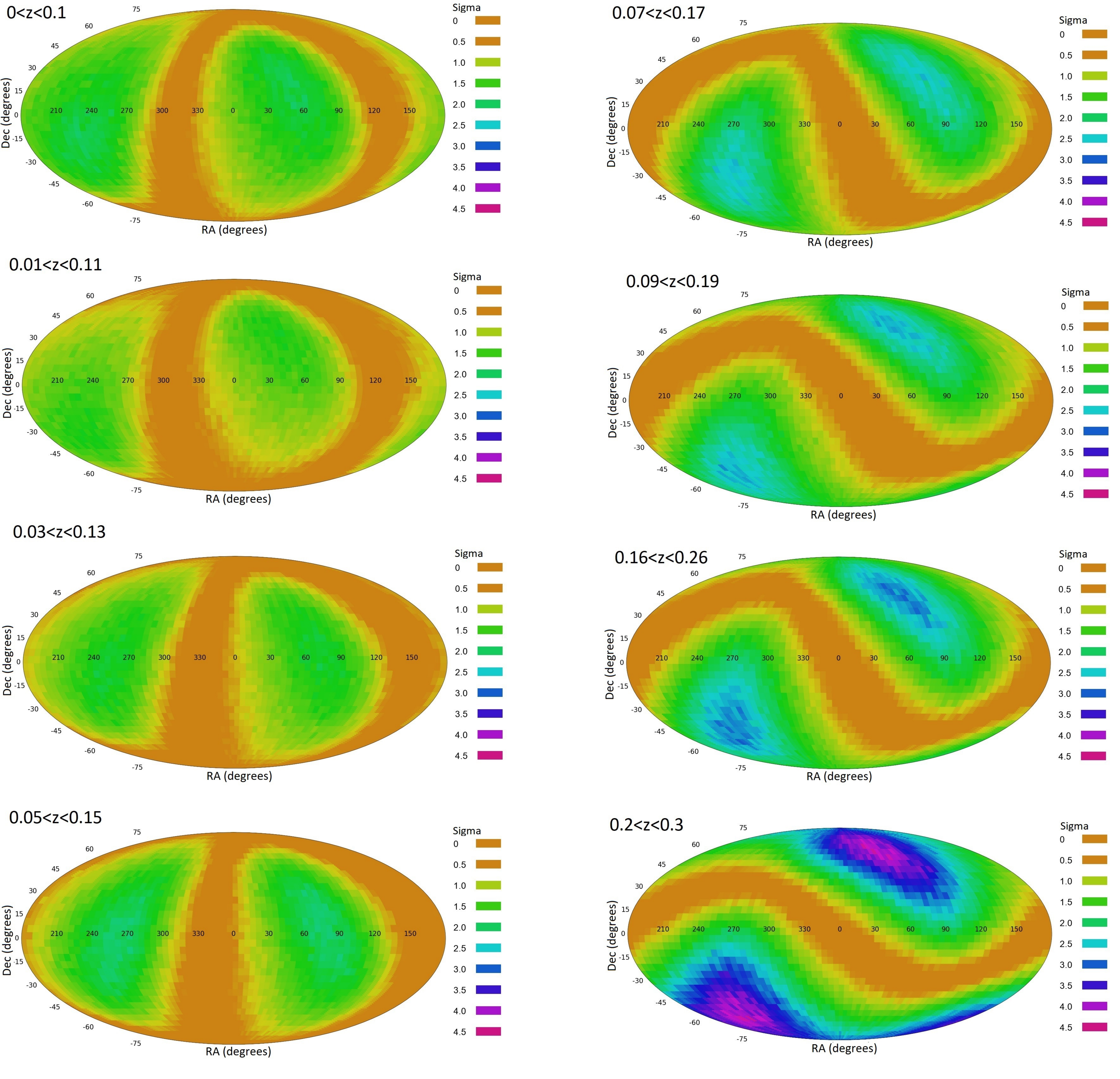}
\caption{The profile of asymmetry axes in different redshift ranges.}
\label{mosaic}
\end{figure*}

The most likely location of the dipole axis changes with the redshift. That change can be profiled with the SDSS dataset, as all galaxies in that dataset have redshift. For instance, Figure~\ref{mosaic} shows the profiles of the asymmetry in several different redshift ranges of $(x<z<x+0.1)$. The figure shows that the location of the most likely axis changes consistently with the redshift. One immediate explanation to the change in the position of the most likely location of the axis with the redshift is that the axis does not necessarily go through Earth.

\begin{figure}[h]
\centering
\includegraphics[scale=0.55]{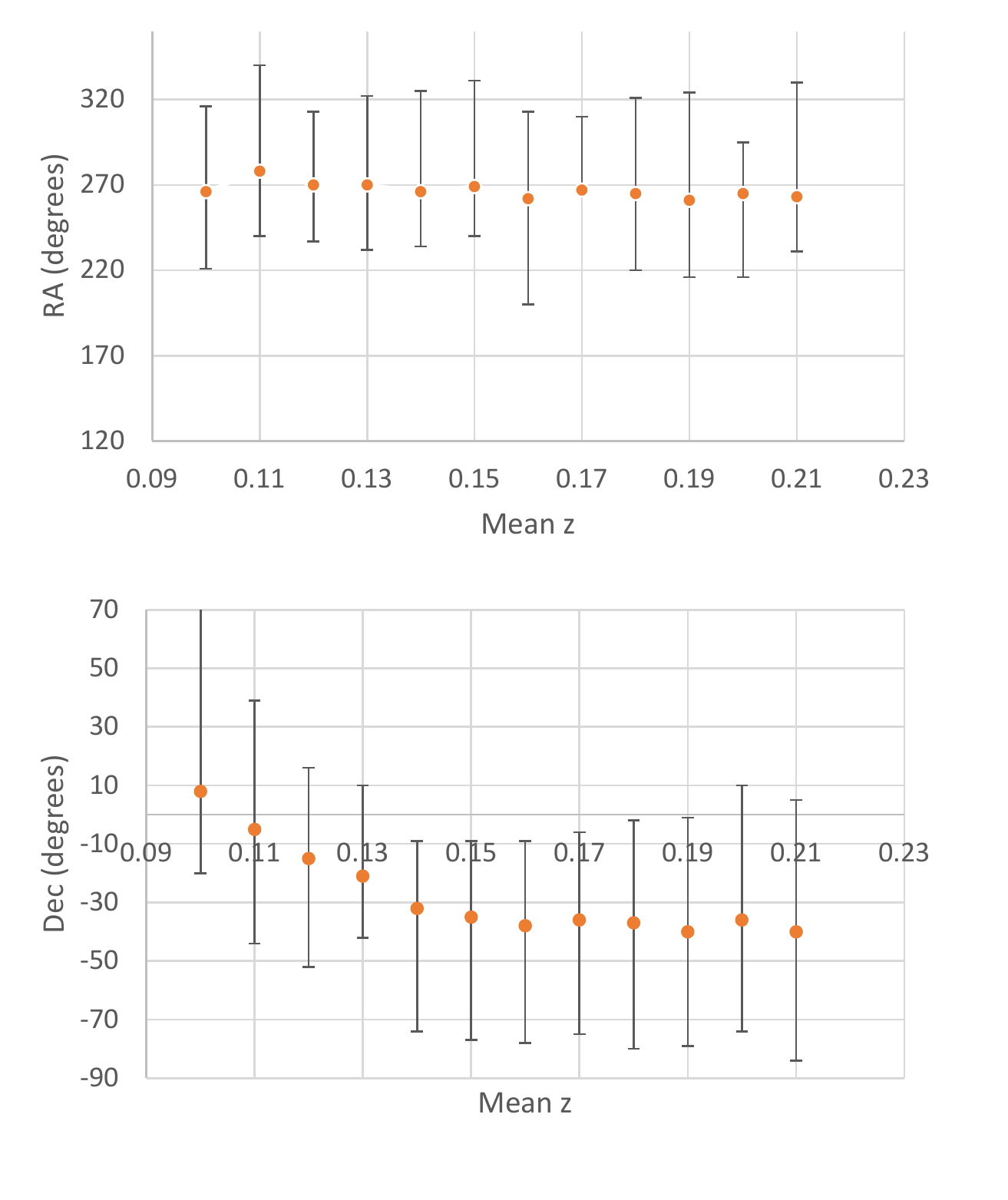}
\caption{The RA (top) and declination of the most likely axis when the redshift range changes. The error bars show the 1$\sigma$ error of the most likely location of the axis.}
\label{ra_dec_correlation}
\end{figure}

\begin{figure}[h]
\centering
\includegraphics[scale=0.55]{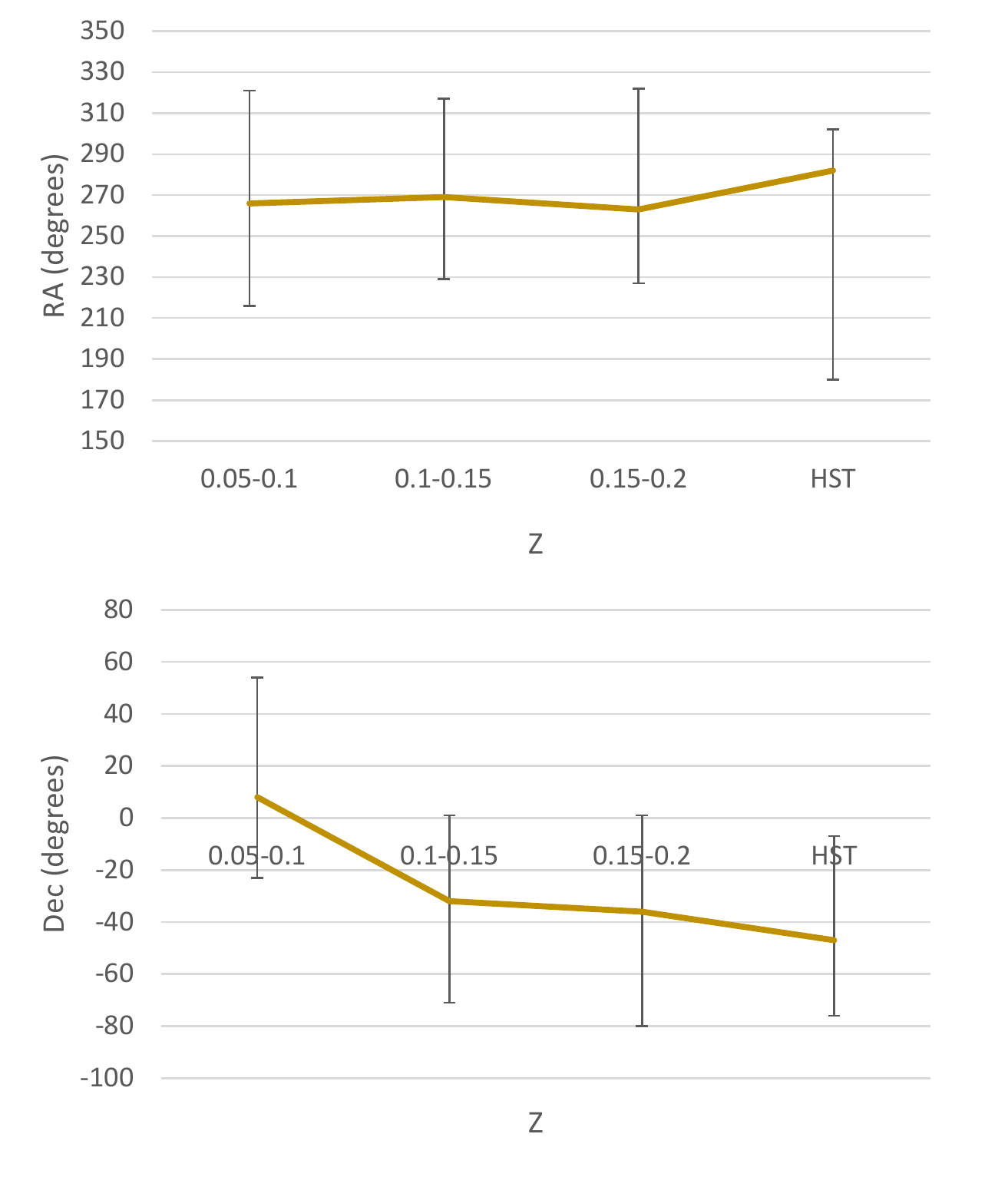}
\caption{The RA and declination of the most likely axis when the galaxies are separated into non-overlapping redshift bins.}
\label{ra_dec_correlation_bins}
\end{figure}

For the analysis, the most likely dipole axis was determined with the SDSS galaxies limited to a 0.1 redshift interval. That was done to all redshift ranges between 0.01$<$z$<$0.11, to 0.16$<$z$<$0.26, where the number of galaxies becomes low. Table~\ref{redshift_ranges} show the location of the most likely axis in different redshift ranges. The error of the RA and declination are the low and high 1$\sigma$ error range of the most likely possible axis. Figure~\ref{ra_dec_correlation} displays the change in the RA and declination of the most likely axis in the different redshift ranges. The {\it x}-axis shows the mean redshift in each redshift range. Figure~\ref{ra_dec_correlation_bins} shows a similar analysis such that the galaxies are separated into non-overlapping redshift bins. The most distant bin is made of HST galaxies.

\begin{table}
\scriptsize
\begin{tabular}{|c|c|c|c|c|}
\hline
        z      & \# galaxies            &  RA                & Dec              & $\sigma$   \\ 
               &                              & (degrees)       & (degrees)      &                  \\  
\hline
0.03-0.13     &   38,938   &   246$\pm$(56,49)      &   -20$\pm$(65,33)       &   1.88            \\				
0.04-0.14     &   38,264   &   258$\pm$(43,52)      &   -5$\pm$(61,30)        &   1.97           \\				
0.05-0.15     &   36,889   &   266$\pm$(50,45)      &   8$\pm$(73,28)        &   2.13             \\				
0.06-0.16     &   34,810   &   278$\pm$(62,38)       &   -5$\pm$(44,39)         &   2.18             \\				
0.07-0.17     &   31,573   &   270$\pm$(43,33)      &   -15$\pm$(31,37)       &   2.42             \\				
0.08-0.18     &   27,500   &   270$\pm$(52,38)      &   -21$\pm$(31,21)       &   2.54             \\				
0.09-0.19     &   23,705   &   266$\pm$(59,32)      &   -32$\pm$(23,42)       &   2.62             \\				
0.1-0.2       &   20,928   &   269$\pm$(62,29)      &   -35$\pm$(26,42)       &   2.50             \\				
0.11-0.21     &   18,018   &   262$\pm$(51,62)      &   -38$\pm$(29,40)       &   2.35             \\				
0.12-0.22     &   14,895   &   267$\pm$(43,55)      &   -36$\pm$(30,39)       &   2.2             \\				
0.13-0.23     &   12,315   &   265$\pm$(56,45)      &   -37$\pm$(35,43)       &   2.4             \\				
0.14-0.24     &   9,957    &   261$\pm$(63,45)      &   -40$\pm$(39,39)       &   2.7             \\				
0.15-0.25     &   8,134    &   265$\pm$(30,49)      &   -36$\pm$(46,38)       &   3.3             \\				
0.16-0.26     &   5,601    &   267$\pm$(67,32)      &   -40$\pm$(45,44)       &   3.2             \\				
\hline
\end{tabular}
\caption{The most likely axis identified in SDSS galaxies when limited to different redshift ranges.}
\label{redshift_ranges}
\end{table}

\begin{figure}[h]
\centering
\includegraphics[scale=0.35]{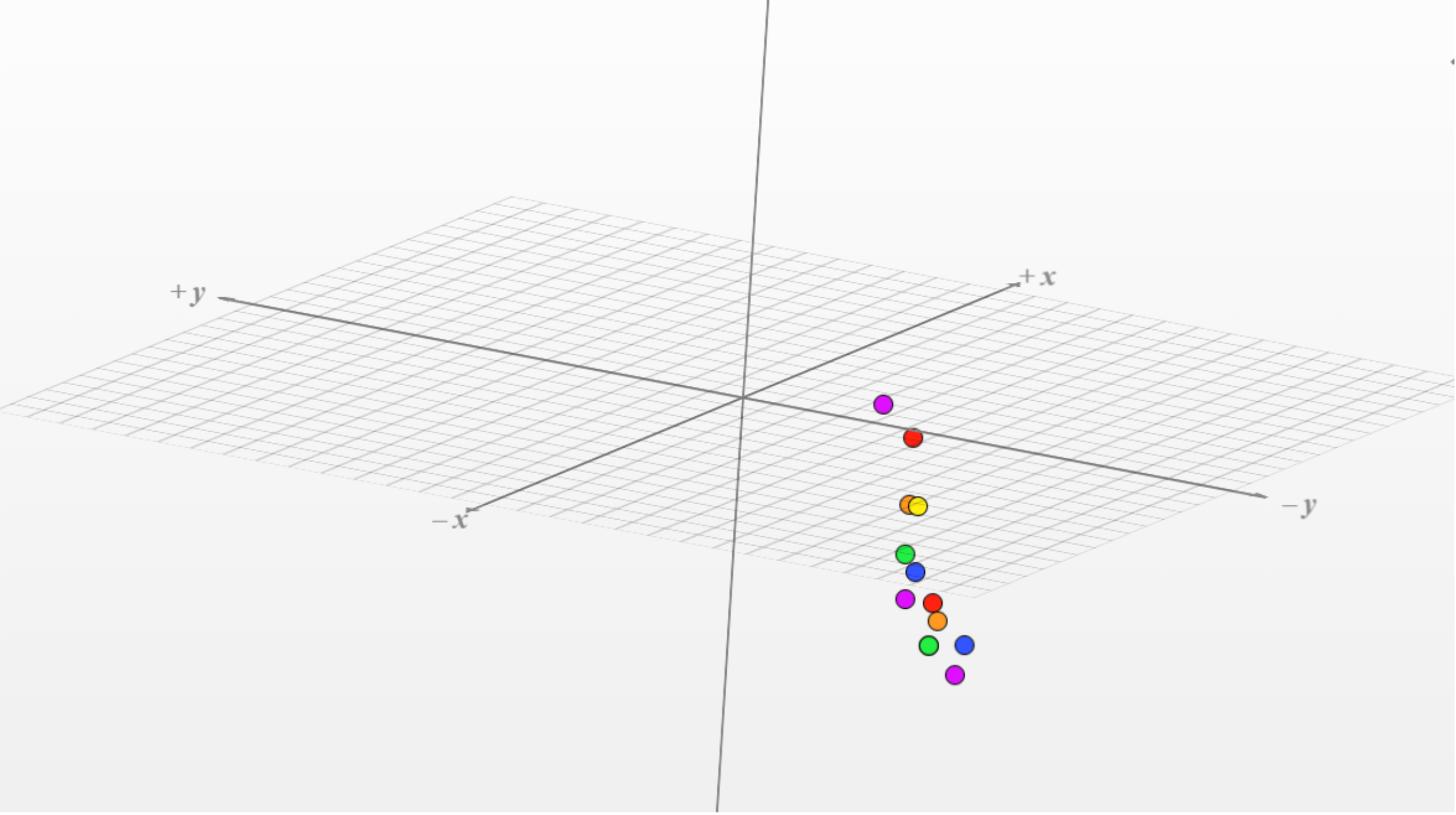}
\caption{3D visualization of the points in Table~\ref{redshift_ranges}.}
\label{3D_axis}
\end{figure}

The figures show that the RA of the most likely axis does not change substantially in the examined redshift ranges. The declination, however, changes consistently with the redshift, until it stabilizes at around z$>$0.15. Figure~\ref{3D_axis} visualizes the most likely axis points in a 3D space such that the distance {\it d} is determined by converting the mean redshift of the galaxies in each redshift range to the distance, measured in Mpc. The 3D transformation is then performed simply by Equation~\ref{transformation}. As expected, the points are aligned in a manner that forms a 3D axis.

\begin{multline}
x=\cos(\alpha)\cdot d \cdot \cos(\delta) \\
y=\sin(\alpha)\cdot d \cdot \cos(\delta) \\
z=d \cdot sin(\delta) \\
\label{transformation}
\end{multline}

To approximate the 3D direction of the axis, a simple 3D analysis between two points {\it a} and {\it b} in a 3D space can be used as shown by Equation~\ref{az_alt_from_3D}. The two points from which the direction of the axis was determined are the two most distant points from each other, which are also to two points with the lowest and highest redshift shown in Table~\ref{redshift_ranges}. The closet point to Earth shown in Table~\ref{redshift_ranges} is at $(\alpha=266^o,\delta=10^o)$, and 310 Mpc away (z$\simeq$0.07). An observer in that point would see an axis in $(\alpha=268^o,\delta=-29^o)$, and the other end of that axis in $(\alpha=88^o,\delta=29^o)$.

\begin{multline}
\alpha = atan2(y_a-y_b, \sqrt{(x_a-x_b)^2 + (z_a-z_b)^2})  \\
\delta = atan2(-(x_a-x_b),-(z_a-z_b))   \\
\label{az_alt_from_3D} 
\end{multline}



The analysis shown in Table~\ref{redshift_ranges} and Figure~\ref{mosaic} shows weak signal in $z<0.1$. The $z<0.1$ range might be dominated by galaxies from the local neighborhood, and therefore the presence or absence of alignment of galaxy angular momentum in that redshift range might not provide an observation that can be considered of cosmological scale. Galaxies within clusters are aligned \citep{tovmassian2021problem}, and galaxies inside the local filament might be aligned in their spin direction \citep{tempel2013galaxy,pahwa2016alignment,antipova2021orientation}. Therefore a possible cosmological-scale alignment in spin direction might be saturated by alignment inside the Galaxy local neighborhood. 

\begin{figure}[h]
\centering
\includegraphics[scale=0.26]{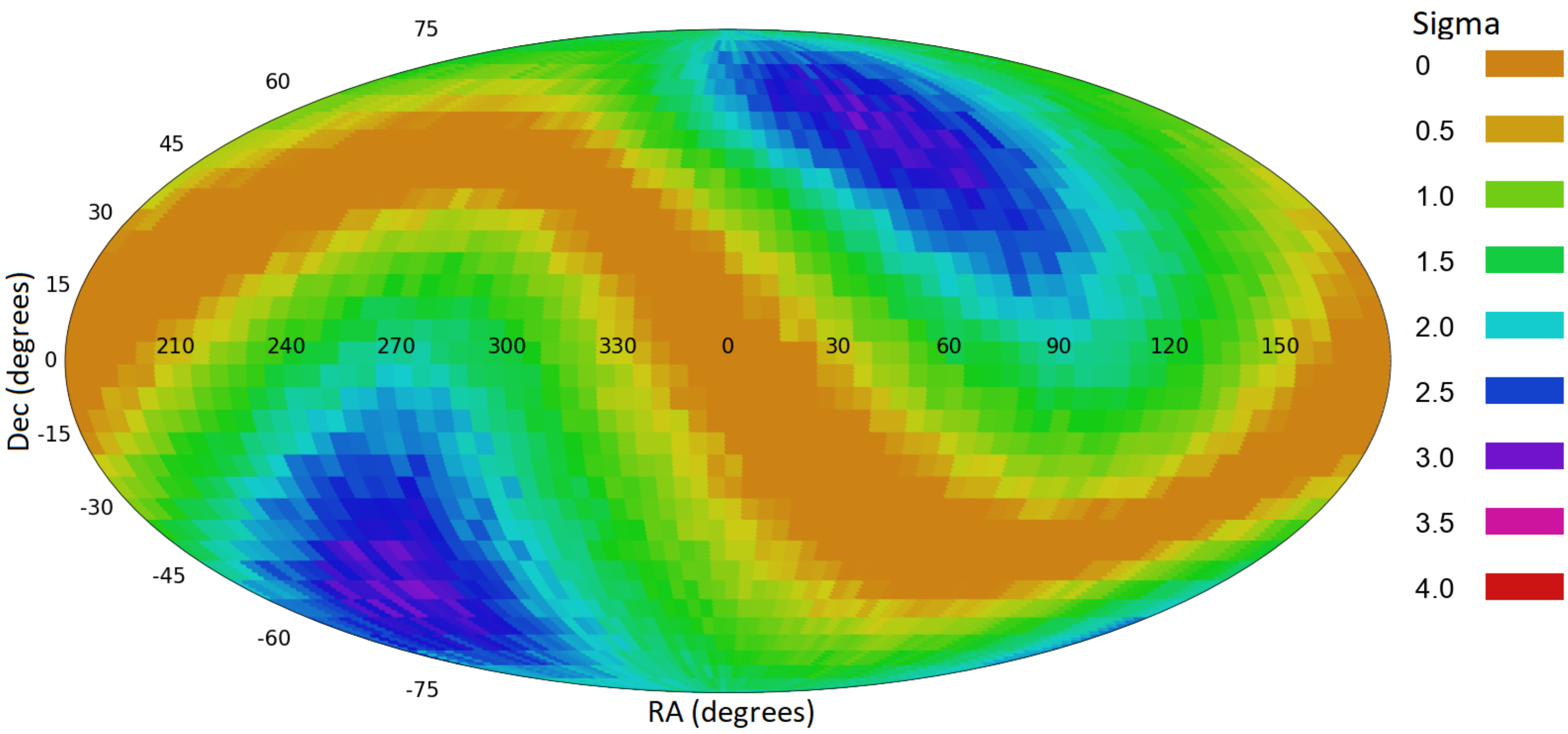}
\caption{Probability of a dipole axis in different $(\alpha,\delta)$ in SDSS galaxies limited to $z>0.1$.}
\label{sdss_z01}
\end{figure}

The SDSS galaxies have spectra, and therefore allow to analyze higher redshift range by excluding galaxies with low redshift. That analysis can also be done by using the HST galaxies. Figure~\ref{sdss_hst} shows that both telescopes have similar profiles when the redshift is greater than 0.15. Assuming that $z>0.1$ is considered outside of any local filament, Figure~\ref{sdss_z01} shows the distribution of the spin directions of SDSS galaxies such that $z>0.1$, meaning that all galaxies used are outside of the local neighborhood.  The figure shows maximum statistical significance of $\sim3.2\sigma$ at $(\alpha=49^{+66^o}_{ -42^o} ,\delta=56^{+28^o}_{-34^o})$. 


That provides indication that the signal can be of cosmological scale, and not necessarily a feature of the local neighborhood. The RA of the most likely axis does not change significantly when $z>0.1$, making it difficult to make conclusions about the impact of the change of the RA. The declination changes consistently until $z\simeq0.15$, where it is stabilizes. The change outside of the local neighborhood might provide a certain indication of cosmological scale axis that does not go directly through Earth, but further research with larger datasets of high-redshift galaxies will be required to study it further. Instruments such as the Dark Energy Spectroscopic Instrument (DESI), when become available, can be combined with optical data such as the DECam data used here to allow further analysis.

Another interesting observation, that can also be seen in Table~\ref{redshift_ranges}, is that the signal increases as the redshift gets stronger. The observed trend of increasing signal with higher redshift is in agreement with the alignment in the polarization of radio sources, that also gets stronger at higher redshifts \citep{tiwari2013polarization}.

\section{Error analysis and ``sanity check''}
\label{error}

Naturally, it is important to ensure that the results are not driven by an error. The spin direction of a galaxy is a crude binary variable, and there is no atmospheric or other effect that can flip the spin direction of a galaxy as seen from Earth. While it is difficult to think of an error in the telescope system that can lead to the observation, the fact that four different telescopes show the same profile with a nearly identical profile indicates that it is unlikely that the observation is driven by some unknown error in the telescopes.

Rarely, galaxies can be counter-winding \citep{grouchy2008counter}, and their actual spin direction is not aligned with the curve of the arms. However, these case are rare, and in any case counter-winding galaxies are equally distributed between galaxies that spin clockwise and galaxies that spin counterclockwise.

The algorithm used for annotating the galaxies \citep{shamir2011ganalyzer} is fully symmetric. It works by clear and defined rules that are completely symmetric, and the algorithm is not based on pattern recognition, machine learning, or deep learning methods that rely on data-driven non-intuitive rules. Therefore, inaccurate annotation is expected to impact both clockwise and counterclockwise galaxies. That was also tested empirically by mirroring the galaxy images \citep{shamir2021particles}. In all datasets, mirroring the galaxy images led to exactly inverse results \citep{shamir2012handedness,shamir2020patterns,shamir2020pasa,shamir2021large}. A more detailed explanation about mirroring the galaxy images is available in \citep{shamir2021particles}.

Also, a bias in the annotation algorithm is expected to lead to similar bias in all parts of the sky, rather than ``flip'' in opposite hemispheres. The observations in all datasets from all four telescopes show inverse asymmetry in opposite hemispheres. Since each galaxy image is analyzed separately, if a consistent error in the annotation algorithm existed, it would have led to the same asymmetry in all parts of the sky, but not to opposite asymmetry in opposite hemispheres.

If the galaxy annotation algorithm had a certain error in the annotation of the galaxies, the asymmetry {\it A} can be defined by Equation~\ref{asymmetry}.
\begin{equation}
A=\frac{(N_{cw}+E_{cw})-(N_{ccw}+E_{ccw})}{N_{cw}+E_{cw}+N_{ccw}+E_{ccw}},
\label{asymmetry}
\end{equation}
where $E_{cw}$ is the number of Z galaxies incorrectly annotated as S galaxies, and $E_{ccw}$ is the number of S galaxies incorrectly annotated as Z galaxies. Because the algorithm is symmetric, the number of S galaxies incorrectly annotated as Z is expected to be roughly the same as the number of Z galaxies missclassified as S galaxies, and therefore $E_{cw} \simeq E_{ccw}$ \citep{shamir2021particles}. Therefore, the asymmetry {\it A} can be defined by Equation~\ref{asymmetry2}.

\begin{equation}
A=\frac{N_{cw}-N_{ccw}}{N_{cw}+E_{cw}+N_{ccw}+E_{ccw}}
\label{asymmetry2}
\end{equation}

Since $E_{cw}$ and $E_{ccw}$ cannot be negative, a higher rate of incorrectly annotated galaxies is expected to make {\it A} lower. Therefore, when the algorithm is symmetric, incorrect annotation of the galaxies is not expected to lead to asymmetry, and can only make the asymmetry weaker rather than stronger. That has also been tested empirically and showed that error makes the signal weaker rather than stronger, unless the error is asymmetric \citep{shamir2021particles}. For that reason, algorithms that are based on pattern recognition, machine learning, or deep learning cannot be used for this purpose, as they are based on complex data-driven rules that they symmetricity is very difficult to verify.

An empirical experiment \citep{shamir2021particles} of intentionally annotating some of the galaxies incorrectly showed that even when an error is added, the results do not change significantly even when as many as 25\% of the galaxies are assigned with incorrect spin directions, as long as the error is added to both clockwise and counterclockwise galaxies \citep{shamir2021particles}. But if the error is added in an asymmetric manner, even a small asymmetry of 2\% leads to a very strong axis that peaks exactly at the celestial pole \citep{shamir2021particles}. 

It should be also mentioned that the HST galaxies were annotated manually, with no intervention of any algorithm \citep{shamir2020pasa}. Other datasets that were prepared manually were used in \citep{shamir2016asymmetry,shamir2017large} The asymmetry profile is consistent regardless of the method of annotation, further showing that a computer error is not a likely explanation to the asymmetry.

\subsection{Analysis of the distribution of the galaxy population}
\label{uneven_distribution}

Obviously, no Earth-based telescope can cover the entire sky, and the footprint of the sky covered by any Earth-based telescope is merely a portion of the entire sky. As discussed in Section~\ref{galaxy_distribution}, the galaxies in the sky surveys are not distributed uniformly inside their footprints. Such non-uniform distribution can in some cases lead to a shift in the location of the most likely axis. For instance, \cite{tiwari2019galaxy} showed that the TGSS ADR1 catalog of radio sources might not be suitable for observing large-scale or moderate scale anisotropy also due to non-uniform distribution of the sources throughout the sky.

What makes the probe of spin direction distribution of spiral galaxies different from other probes is that the spin direction is a comparative measurement, rather than an absolute measurement. Absolute measurements such as CMB or the number of radio sources can be impacted by the distribution of the measurements, as well as by other effects such as Milky Way obstruction. For instance, if the Milky Way obstructs some of the sky, the change in the temperature in that part of the sky can lead to certain anisotropy. When counting radio sources, an anisotropy can be driven by differences in the coverage of different parts of the sky \cite{tiwari2019galaxy}.

The ratio between galaxies that spin in opposite directions is a comparative scale, meaning that it is not determined by the number of galaxies in a certain field, but by the difference between the number of galaxies spinning clockwise and the number of galaxies spinning counterclockwise in that field. In each given field, the number of clockwise galaxies is balanced by a roughly similar number of counterclockwise galaxies. The cosine dependence analysis described in Section 3 identifies the best axis that fits the distribution of the clockwise and counterclockwise spiral galaxies. Because the measurement is relative, it is expected to be less sensitive to the distribution of the absolute number of galaxies, as each field that used in the analysis has a roughly similar number of galaxies that spin in each direction. Therefore, four different telescopes with four different footprints provide very similar location of the most likely axis.

To test the effect of the non-uniform distribution of the galaxies empirically, the galaxies from the four telescopes were combined into a single dataset. In order not to use the same galaxy more than once, galaxies that appear in two or more telescopes were removed, leading to a dataset of 935,155 galaxies. To create a dataset of galaxies that are distributed uniformly within the footprint, the 935,155 galaxies were added in random order into a new dataset such that before adding a galaxy to the dataset, the number of galaxies with 5$^o$ radius from that galaxy was counted. If 500 or more galaxies are present within a distance of 5$^o$, the galaxy is not added to the dataset. Otherwise, the galaxy is added. The algorithm for creating that dataset {\it U} from the dataset {\it D} of 935,155 galaxies can be summarized as follows: \newline \newline
1. U $\leftarrow$ \{\} \newline
2. For each galaxy d in D \newline
3.    \hspace{0.5cm} galaxy\_counter $\leftarrow$ 0 \newline
4.    \hspace{0.5cm} For each galaxy u in U \newline
5.    \hspace{1cm}   if distance(d,u)$<5^o$ \newline
6.     \hspace{1.5cm}          galaxy\_counter $\leftarrow$ galaxy\_counter +1 \newline
7.   \hspace{0.5cm} if galaxy\_counter$<$500 \newline
8.   \hspace{1cm}    add galaxy d to U \newline

\begin{figure}[h]
\centering
\includegraphics[scale=0.27]{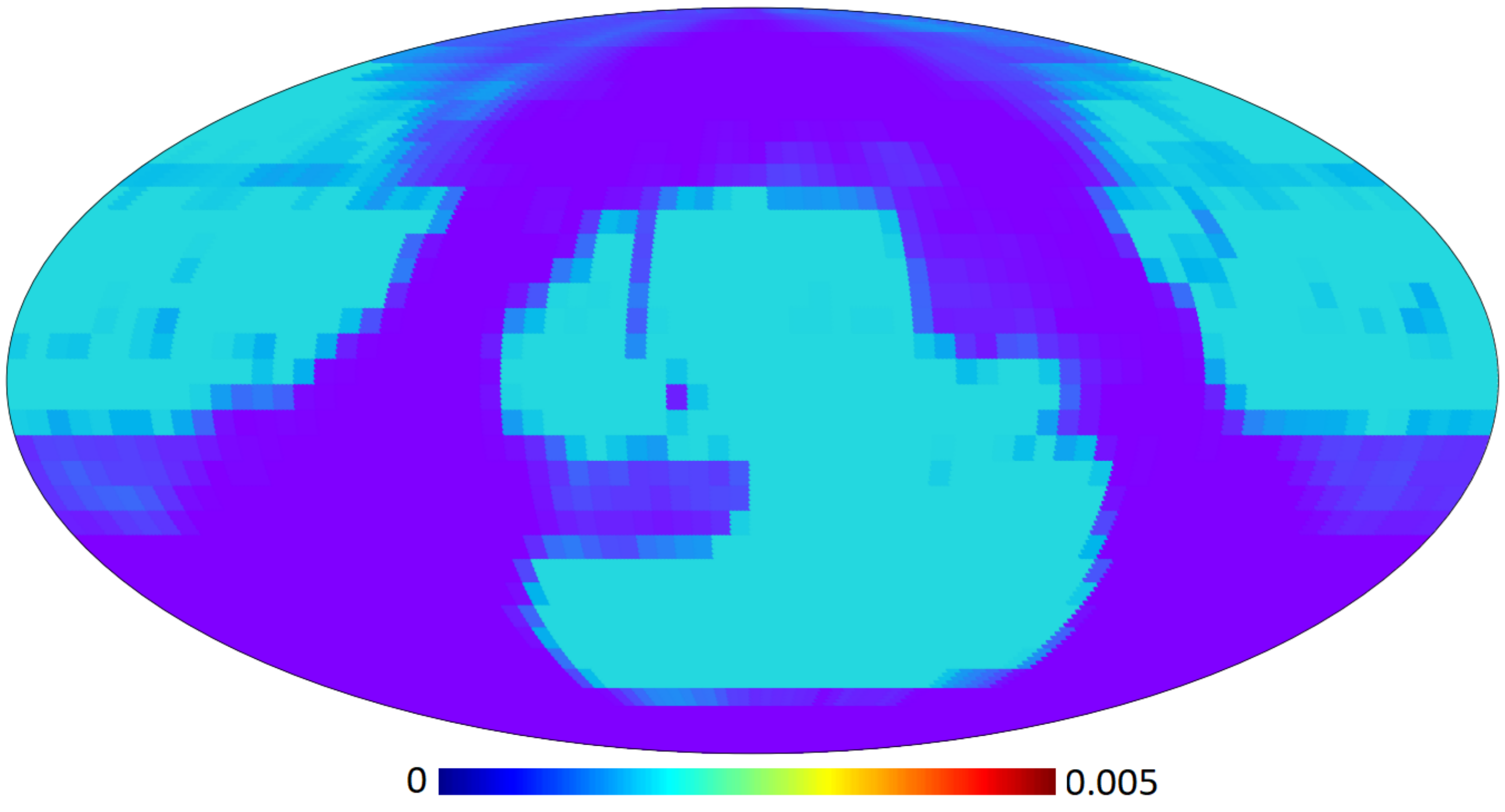}
\caption{The density of the galaxy population in the combined dataset with uniform distribution of the galaxies inside the combined footprint. The galaxy population density in each point of the sky is determined by the number of galaxies within 5$^o$ from that point, divided by the total number of galaxies.}
\label{combined_500_population}
\end{figure}

\begin{figure}[h]
\centering
\includegraphics[scale=0.26]{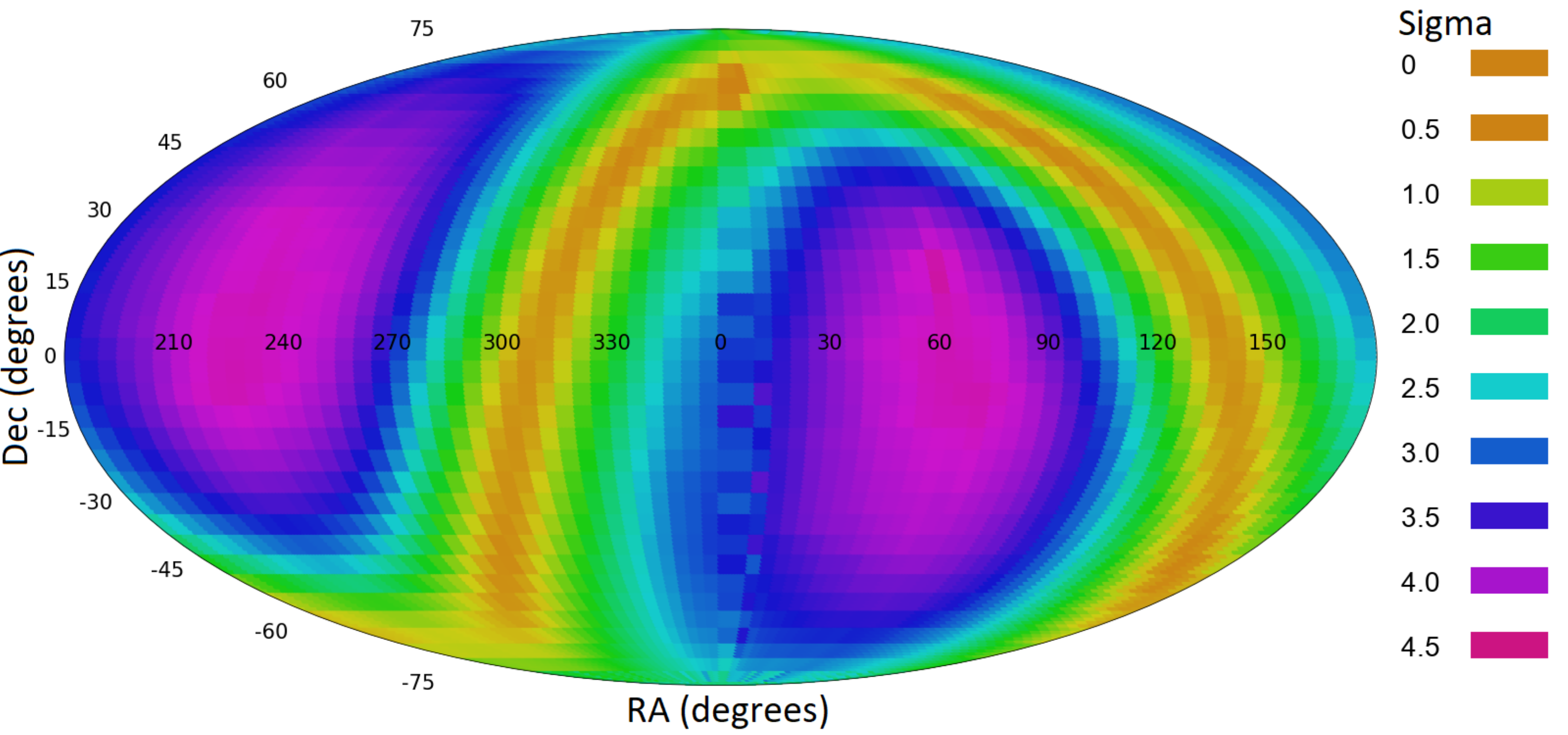}
\caption{The probability of the most likely axis from each $(\alpha,\delta)$ combinations. The number of galaxies is similar in the different populated parts of the sky.}
\label{combined}
\end{figure}

Then, isolated galaxies in dataset {\it U} that do not have 500 galaxies within 5$^o$ are removed from {\it U}. When the galaxies in {\it D} are iterated in random order, that process leads to a dataset {\it U} of 193,368 galaxies that are distributed uniformly inside the combined footprint. Figure~\ref{combined_500_population} shows the distribution of the galaxy population in different parts of the sky. The dataset is smaller than the DECam dataset, but spreads much more uniformly in the sky compared to the other sky surveys. Figure~\ref{combined} shows the probability for a dipole axis in each $(\alpha,\delta)$ combination. The most likely axis is at $(\alpha=63^o,delta=-16^o)$.

As the figure shows, when the distribution of the galaxies is uniform, and no certain parts of the sky are overpopulated or underpopulated, the profile of the asymmetry is in agreement with the profile determined by DECam or Pan-STARRS when the galaxy population is not normalized. Considering the ratio in each part of the sky as temperature, the HEALPix library \citep{zonca2019healpy} can also be used to profile the dipole, showing a dipole at $(\alpha=45^o,\delta=-23^o)$.

\begin{figure}[h]
\centering
\includegraphics[scale=0.25]{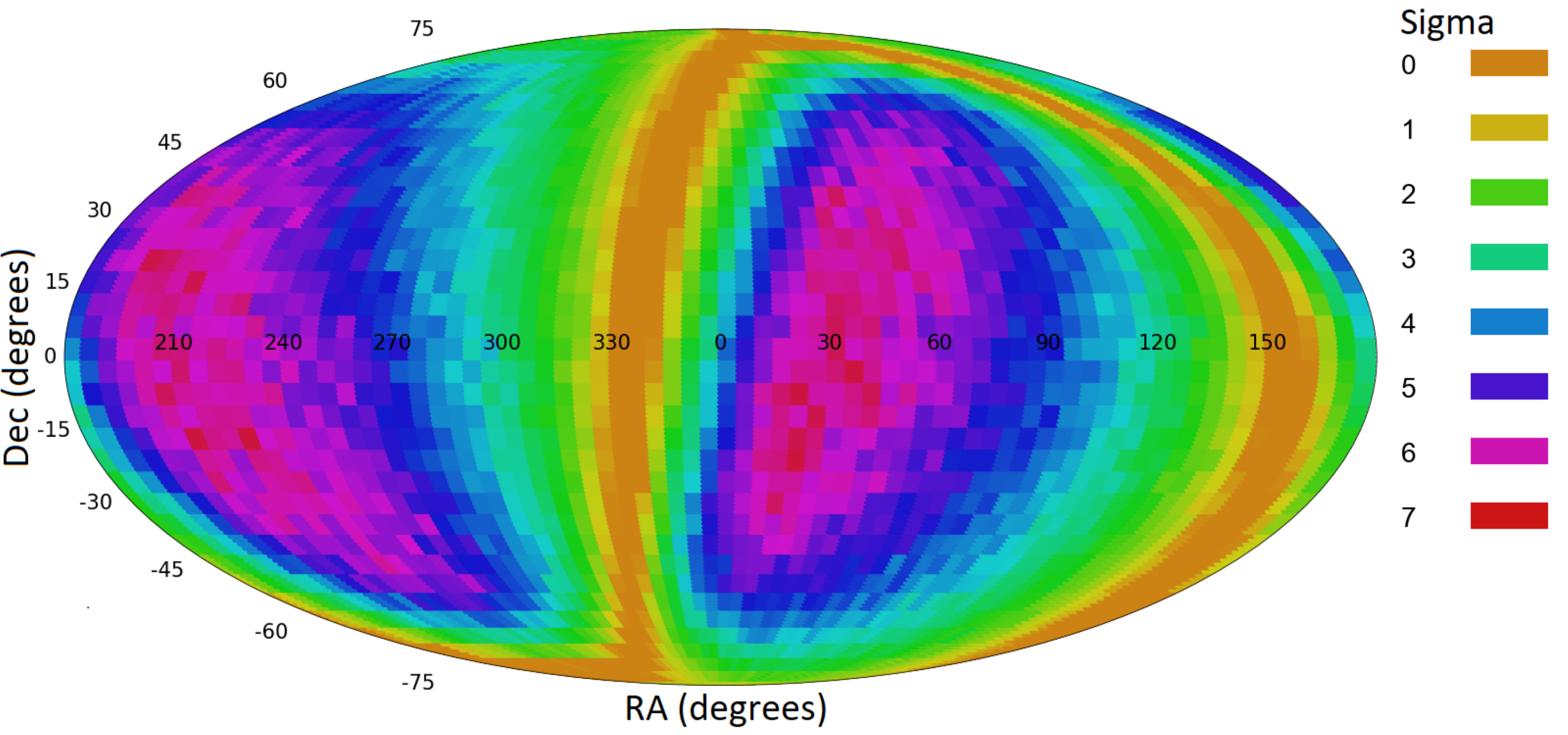}
\caption{The probability of the most likely axis from each $(\alpha,\delta)$ combinations. The parts of the sky used in the analysis are of similar size in exact opposite hemispheres.}
\label{combined_dual}
\end{figure}

As mentioned above, there is no Earth-based telescope that can cover the entire sky. Even when combining the footprints of all telescopes, substantial parts of the sky are not covered. The relative measurement used here is expected to lead to the same profile of best cosine fitness regardless of the parts of the sky that are covered, given that the sky coverage is sufficiently large to allow fitness. However, the non-uniform distribution of the parts of the sky that need to be covered should be tested empirically. For that purpose, an experiment was performed such that only two parts of the sky were used, in two exactly opposite hemispheres. The two parts of the sky were $(0^o < \alpha < 60^o, -45^o < \delta < 0^o)$, and the exact same part of the sky in the opposite hemisphere $(180^o < \alpha < 240^o, 0^o < \delta < 45^o)$. These are two exact opposite parts of the sky, and therefore an asymmetry profile is not expected to be driven by non-uniform distribution of the galaxies in the sky. That dataset included 282,252 galaxies. Figure~\ref{combined_dual} show the results, with the most likely axis peaking with 5.4$\sigma$ at $(\alpha=41^o,\delta=-12^o)$, well within the 1$\sigma$ error of the analysis when using the entire dataset.

\section{Conclusions}
\label{conclusions}

Multiple sky surveys have shown certain patterns of asymmetry between clockwise and counterclockwise galaxies. These sky surveys include SDSS \citep{shamir2012handedness,shamir2020patterns}, Pan-STARRS \citep{shamir2020patterns}, DECam \citep{shamir2021large}, and HST \citep{shamir2020pasa}. These observations include different ways of annotating the galaxies, observations from both the the Northern and Southern hemispheres, and observations from ground-based and space-based instruments. All observations are in very good agreement. As multiple probes have shown disagreement with the cosmological isotropy assumption, the contention that the Universe is oriented around an axis is in agreement with several existing cosmological theories such as black hole cosmology, elliptodial universe, rotating universe, and other cosmological models that provide an alternative to the standard model.

One of the possible weaknesses of the analysis is that the footprint does not cover the entire sky. As no Earth-based sky survey can cover the entire sky, an experiment that provides 100\% sky coverage is not possible. The use of a comparative measurement and not an absolute measurement makes the analysis more robust to cases of non-uniform distribution. Several empirical experiments were also performed to analyze the impact of the asymmetric shape of the footprint, showing consistent results, although no experiment that covers the entire sky can be done, as the sky surveys do not provide full sky coverage.

The observations show that the location of the most likely axis changes with the redshift at $z>0.1$. That can be viewed as an indication that if such axis exists, it does not necessarily go directly through Earth. The general location of the axis at $(\alpha=\sim260^o,\delta=\sim-40^o)$ is roughly aligned with the axis formed by the isotropy of the accelerating cosmological expansion rate at $(\alpha=223^o,\delta=-76.2^o)$ as explained in \citep{perivolaropoulos2014large}. The Shapely supercluster \citep{sheth2011unusual} is also not far at $(\alpha=202^o,\delta=-30^o)$.

As the evidence of cosmological-scale anisotropy are accumulating, theories that provide alternatives or expansions to the standard cosmology model are being proposed \citep{arun2017dark,nadathur2012integrated,narlikar2007cosmology,perivolaropoulos2014large,bull2016beyond,lopez2017tests,scott2018standard,tatum2018flat,neves2020proposal,nojiri2019modified,lopez2017tests,freedman2017cosmology,fiscaletti2018towards,tatum2018flat,guo2019can,li2021meta}. Such cosmological-scale alignment between galaxies can also be related to alternative gravitational theories that shift from the Newtonian model \citep{bekenstein1984does,milgrom2009bimetric,deser2007new,alkacc2018holographic,sivaram2020mond,sivaram2021hubble,sivaram2021non}, where gravity span is far greater than assumed by the existing models of gravity \citep{amendola2020measuring,falcon2021large}. Such gravity models have impact on cosmology \citep{sivaram1994some}, and can also explain other anomalies that disagree with the standard model such as the Keenan–Barger–Cowie (KBC) void \citep{haslbauer2020kbc} and the El Gordo (ACT-CL J0102-4915) massive galaxy cluster \citep{asencio2020massive}. It is therefore important to continue to explore different probes and different observations that can provide more information about the structure of the Universe. Clearly, further research will be required to fully profile the nature of the asymmetry.

\section*{Acknowledgment}

I would like to thank the knowledgeable anonymous for the insightful comments. This study was supported in part by NSF grants AST-1903823 and IIS-1546079. 

\bibliographystyle{apalike}
\bibliography{main}

\end{document}